\documentclass{agujournal2019}
\usepackage{url} 
\usepackage{float}
\usepackage{lineno}
\usepackage{soul}
\usepackage{subcaption}
\usepackage{adjustbox}
\usepackage{amsmath}
\usepackage{commath}
\usepackage{bm}

\draftfalse

\journalname{JGR: Machine Learning and Computation}

\newcommand{\REFA}[1]{{#1}}
\newcommand{\LUCAS}[1]{{#1}} 
\newcommand{\PATRICIO}[1]{\textcolor[HTML]{D1495B}{\textbf{[Patricio: #1]}}} 

\begin{document}

\title{Estimating bottom topography in shallow water flows}

\authors{L. Pancotto \affil{1,2,4}, and P. Clark Di Leoni\affil{2,3}}

\affiliation{1}{Departamento de Física, FCEN, UBA, Buenos Aires, Argentina}
\affiliation{2}{Universidad de San Andrés, Buenos Aires, Argentina}
\affiliation{3}{CONICET, Argentina}
\affiliation{4}{Department of Mechanical Engineering, Johns Hopkins University, Baltimore, MD 21218, USA}

\correspondingauthor{Patricio Clark Di Leoni}{pclarkdileoni@udesa.edu.ar}

\begin{keypoints}
\item We develop two methods estimate bottom topography in a shallow water flow using surface measurements.
\item One method is based on physics-informed neural networks while the other is
    based on the adjoint state method.
\item \REFA{We tested against data sparsity and corruption in 1D and 2D
        configurations. Both methods produce promising results.}
\end{keypoints}

\begin{abstract}
We present two methods to estimate bottom topography in a shallow water flow using only surface deformation measurements. One is based on Physics-Informed Neural Networks (PINNs) and the other on the Adjoint State Method. We test both methods using synthetic data \REFA{in 1D and 2D cases}. Both are able to successfully reconstruct not only the bottom topography but also the surface velocity. Both also show robustness against noise and data sparsity up to reasonable levels. 
\end{abstract}

\section*{Plain Language Summary}
Determining what lies beneath the ocean surface is an open challenge. We present two techniques that can recover the shape of the bottom topography out of surface measurements. These techniques provide new ways to probe the depths of the oceans.

\section{Introduction}

The seafloor has always been a subject of speculation, captivating both imagination and intellect. What is that world lying beneath the bottom of our oceans like? More than $70\%$ of the surface of our planet is covered by ocean waters, which amounts to approximately $362$ million km$^2$ of the total surface area \cite{eakins}. Beyond the natural curiosity that arises when considering the shape of those vast landscapes, bathymetry plays a fundamental role in the countless activities and disciplines involving the atmosphere and the ocean. Understanding the shape of the seafloor is essential for modeling and comprehending ocean circulation \cite{Gulf-Stream-Dynamics-Southeastern-US-Sea}, which in turn plays an important role in the predictive capacity of climate models for global phenomena \cite{Indonesian-Throughflowon-ENSO-Dynamics}. Many current studies highlight the need to achieve a global mapping of the ocean floor, as well as the numerous difficulties such an endeavor entails  \cite{bathymetry, quest-to-map-oceans}. Bathymetry directly affects the evaluation of geophysical risks of various kinds, such as landslides \cite{geohazards}, or the formation and propagation of tsunamis \cite{tsunamis}. The topology of our ocean floors, after all, constitutes an essential boundary condition in the titanic problem of understanding the flows that govern the surface of our planet.

It is a fact that currently only a small portion of the area occupied by the oceans is charted \cite{airline-flights-unmapped-oceans}. The vast task of documenting these topographies presents numerous obstacles. The earliest records of bathymetric measurements date back to Ancient Egypt, around 3000 years ago \cite{sounding-to-pole-beam}. Bathymetry techniques available today include some that allow measurements from satellites—such as Satellite-Derived Bathymetry (SDB) \cite{satellite-remote-sensing}—or from aircraft, like Light Detection and Ranging (LIDAR) \cite{lidar}. However, these are sensitive to water transparency and are generally used in shallow waters, where it is also riskier and more difficult to take measurements from vessels \REFA{\cite{bathymetry,lidar,satellite-remote-sensing}}. Other strategies based on acoustic waves—such as Singlebeam Echo-Sounders (SBES) \cite{mayer-mapping-visualization} and Multibeam Echo-Sounders (MBES) \cite{Glenn_multibeam}—require seafaring, and involve higher costs and risks due to the need to navigate in the areas to be measured. Satellite Altimetry, alternatively, measures the height of the ocean surface from satellites and infers the seafloor topography from the gravitational effects it induces on the surface \cite{satellite-altimetry}. While it allows for relatively fast surveying of the surface, it has a much lower resolution than bathymetry performed directly over the water.

In this work, we focus on studying bathymetry in flows under the Shallow Water (SW) approximation \cite{pedlosky} using data assimilation techniques. Specifically, this work compares the performance of Physics-Informed Neural Networks (PINNs) \cite{kardianakis} and the Adjoint State methodologies (ASM) \cite{variational-algs-meteorology} in the task of inferring the topography of the bottom boundary of numerically simulated shallow water flows, assimilating data from the height field only. 
Flows governed by the SW equations have the advantage of eliminating vertical dependence by considering waves with wavelengths much greater than the fluid depth. Solving the proposed inverse problem then offers a tool for mapping the seafloor topography with surface measurements only. \REFA{Shallow water models arise naturally when considering tidal or coastal flows.} Another interesting example are tsunamis. Modeling them this way is reasonable, as they typically have wavelengths of tens of kilometers and propagate in waters only a few kilometers deep. Far from the coast, the amplitude of these waves is in the order of one meter. The case study presented in \cite{tsunamis} features a wavelength of $20km$ propagating in waters with an average depth of $4km$, with bottom fluctuations of a few hundred meters. 


Physics-Informed Neural Networks (PINNs) are a recently introduced methodology that has shown significant advances \cite{cuomo-machine-learning,cai-pinns-review}. They are based on supervised learning, as in traditional neural networks, but incorporate the constraints expressed by partial differential equations (PDEs) directly into the loss function (as residuals), thus enforcing the training process to satisfy the known physics of the system \cite{kardianakis}. The cost function is minimized subject to both the measurements and the governing physical equations via gradient descent. This allows the inference of quantities that are not directly measured but appear in the equations, as is the case in this work for the bottom contour and velocity field. \REFA{The physics-informed training} also guides the training process to arrive at physically plausible solutions. This methodology offers several advantages — two important ones being the use of automatic differentiation to compute all required derivatives, and the fact that there is no need to define a discrete space–time grid for the problem domain \cite{pinn-pathology}. PINNs have been particularly successful at dealing with inverse problems such as the one presented here. For example, they have been used to reconstruct pressure fields from velocity measurements \cite{clark-reconstructing-turbulent-velocity}, reconstruct velocity fields from temperature measurements \cite{clark_di_leoni_reconstructing_2023}, reconstruct and enhance PIV measurements \cite{cai_flow_2021,hasanuzzaman_enhancement_2022}. \LUCAS{Of particular relevance to the present study is the work presented in \cite{ohara_physics-informed_2024}, where the authors use PINNs to estimate velocity, surface deformation and bathymetry of a shallow river. As we explain below, we go beyond by comparing PINNs to another numerical method, as well as extending the error analysis by performing tests with varying sparseness and noise.}

\begin{figure}[t]
    \centering
    \includegraphics[width=0.8\textwidth]{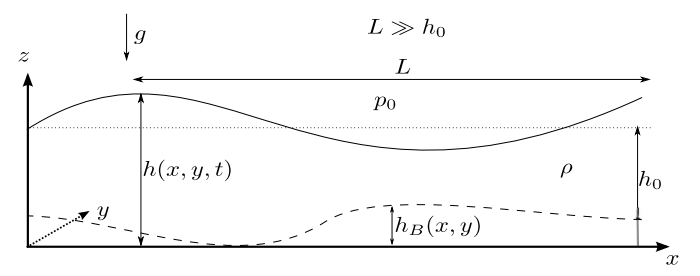}
    \caption{Diagram of the quantities involved in a SW problem. $h(x, y, t)$ is the fluid height field, $h_b(x, y)$ is the bottom topography, \( h_0 \) is the height of the fluid at rest, \(L\) is a characteristic horizontal distance,  \( p_0 \) is the pressure just outside the fluid, \( \rho \) the fluid density, and \( g \) gravity.
 }
    \label{diagrama_swhd}
\end{figure}

\REFA{The Adjoint State Method (ASM) is a variational approach to tackle inverse problems \cite{variational-algs-meteorology,zaki_Turbulence_2025}.} A cost function for the available field measurements is minimized subject to the physical equations of the problem. A Lagrangian is formulated that includes both the cost function from data to be extremized and the differential operators governing the physics. The adjoint problem must then be derived, specific to each cost function, and a set of constraints in the form of differential operators, and finally solved \cite{variational-algs-meteorology, adjoint-state-geophysical}. Solving the adjoint problem has a significantly lower computational cost than solving the problem directly via the Euler-Lagrange equations \cite{photon-tomography}. The methodology was originally developed to perform data assimilation for numerical weather forecasting, but has been extended to other problems, such as reconstructing homogeneous three-dimensional turbulence \cite{li_small-scale_2020}, locating scalar sources in turbulent flows \cite{wang_spatial_2019} and reconstructing various flows \cite{wang_discrete_2019,wang_state_2021}. It has also been shown to work in neural-network generated latent spaces \cite{cleary_Latentspace_2025a}. In particular, \cite{minimal-turbulent-channel-flow} compared both PINNs and a variational approach, finding the variational one to be more accurate and robust than PINNs.
In this work, we present the derivation of the adjoint system of equations to apply it to the bathymetry problem in shallow water (SW), and the results of its application via pseudo spectral numerical integrators. 

The objective of this work is to assess whether these methods are suitable for addressing the inverse problem of determining the bathymetry and velocity field by assimilating sparse surface measurements in SW flows. We compare their performance in cases with different sparsity for surface measurements, and different amplitudes of random noise added to the measurements (for a fixed sparsity). The manuscript is structured as follows. Section~\ref{sec:materials} derives the SW equations, discusses the pseudo spectral scheme utilized for numerical simulations, briefly discusses PINNs, present the theoretical framework we derived for the Adjoint State method in this problem, and introduces the parameters used for the numerical simulations and methods. In~\ref{sec:results}, we compare the results from both methods, in subsection~\ref{sec:noiseless results} for cases with different sparsities in the data assimilated, and in subsection~\ref{sec:noise results} for cases with different amplitudes of noise added to data. Section~\ref{sec:conclusions} presents the conclusions and future work prospects.

\section{Equations and Methods}
\label{sec:materials}

\subsection{Shallow Water equations and problem formulation}

\begin{figure}[t]
     \centering
         \includegraphics[width=0.8\textwidth]{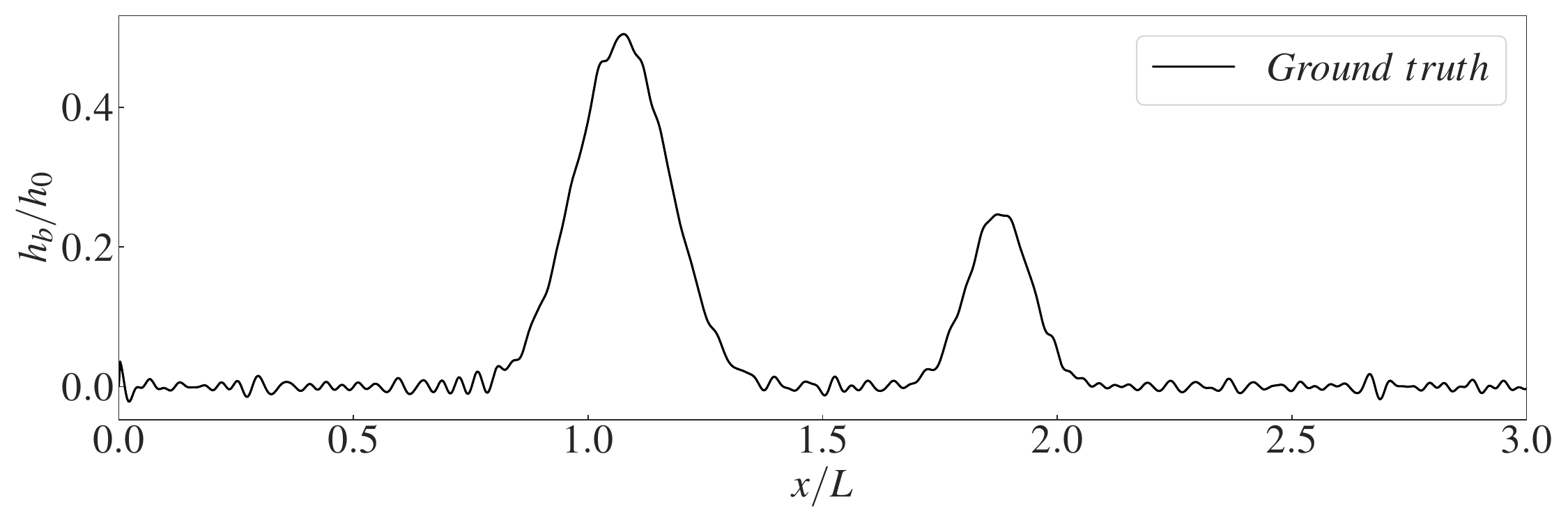}
        \caption{Bottom profile $h_b(x)$ used in the 1D experiments.}
    \label{hb_croquis}
\end{figure}

\begin{figure}[t]
    \centering
         \includegraphics[width=0.8\textwidth]{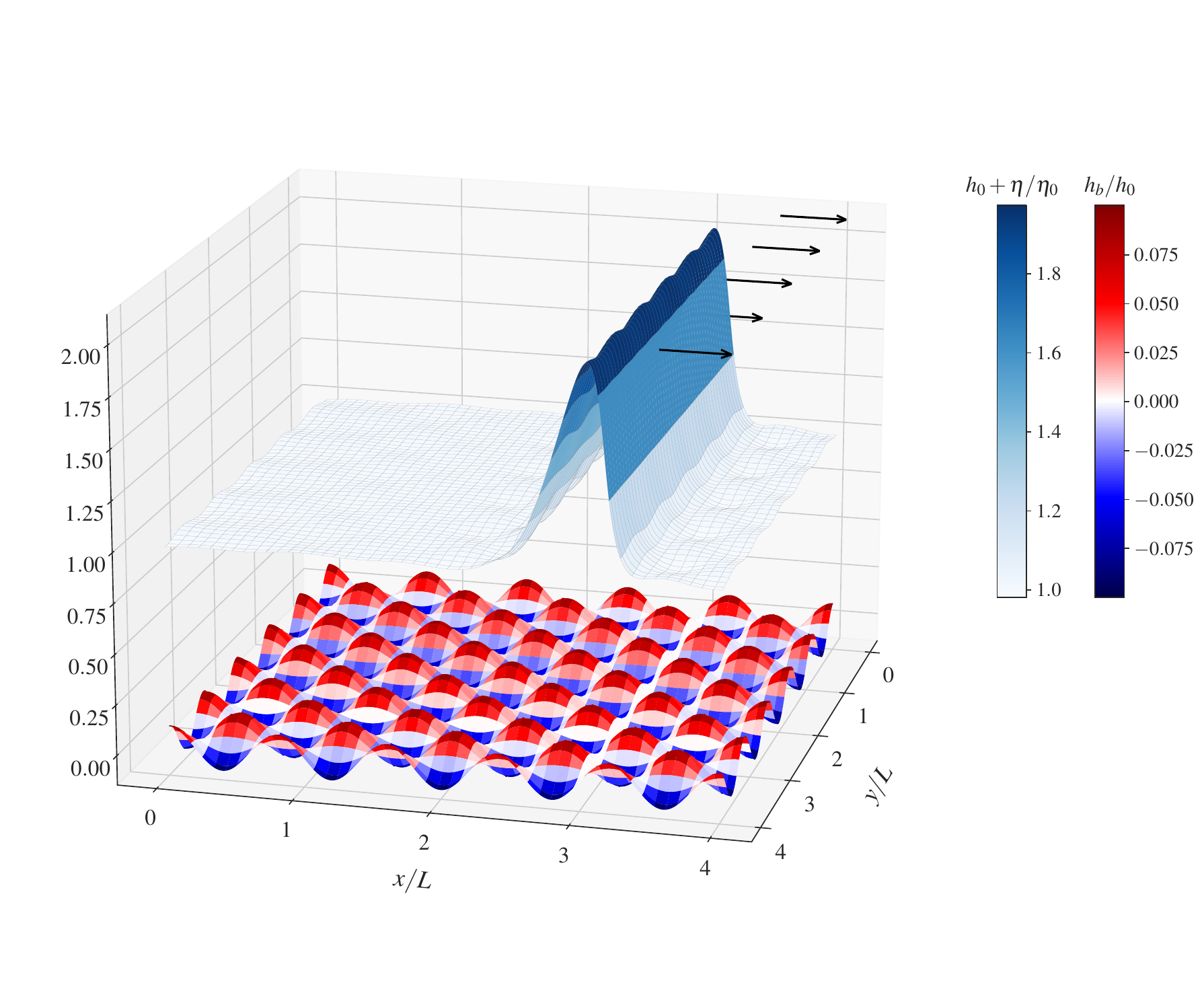}
        \caption{Diagram of 2D flow studied at $t/T = 0.5$. The blue surface depicts ground truth of field $h_0 + \eta/\eta_0$, and the diverging red and blue colored surface depicts ground truth of $h_b/h_0$. The arrows indicated the direction in which the wave is propagating. }
    \label{dg2d_diagram}
\end{figure}

Following the setup shown in Fig.~\ref{diagrama_swhd} and starting from the Euler equations for an incompressible flow, assuming free surface boundary conditions on the surface and no-slip conditions at the bottom, and that the characteristic horizontal length scale $L$ is much larger than the characteristic vertical length scale $h_0$, \REFA{which is the height of the fluid at rest measured from the lowest point}, one can derive the Shallow Water (SW) equations \cite{pedlosky} that describe the dynamics of the free surface

\begin{align}
    \label{SW-u}
        \frac{\partial \bm{u}}{\partial t}
        + (\bm{u}\cdot \bm{\nabla}) \bm{u}
        + g\, \bm{\nabla} h= & 0\\
    \label{SW-h}
        \frac{ \partial h}{\partial t}
        + \bm{\nabla} \cdot \left( \bm{u} (h - h_b) \right) = & 0,
\end{align}
where $\bm{u}=(u, v)$ is the horizontal velocity field on the surface, $h$ is the total fluid height, $h_b$ is the bottom topography, and $g$ is the acceleration of gravity. We also define the surface deformation $\eta = h - h_0$. \LUCAS{We consider both 1D and 2D flows. In the 1D cases we assume translation symmetry in the $y$-direction, thus removing all dependency on $y$ and setting $v=0$.}


Under this model, the problem we try to solve is to estimate $h_b$ given surface measurements $\{h_j\}_{j=1}^N$ at locations $\{\bm{x}_j, t_j\}_{j=1}^N$ \REFA{(with $\bm{x}=x$ in the 1D case and $\bm{x}=(x,y)$ in the 2D case)}. We can cast this task as an optimization problem where the goal is to find $\hat{u}$, $\hat{h}$ and $\hat{h}_b$ such that the cost function

\begin{equation}
\mathcal{L}_d = \frac{1}{N_d} \sum_{j\in \Omega_d}(h_j - \hat{h}_j)^2,
\label{loss_general}
\end{equation}
is minimized, subject to $\hat{u}$, $\hat{h}$ and $\hat{h}_b$ satisfying Eqs.~\eqref{SW-u} and~\eqref{SW-h} and where $\Omega_d = \{\bm{x}_j, t_j; h_j\}^{N_d}_{j=1}$ is the aforementioned set of training data and $\hat{h}_j = \hat{h}(\bm{x}_j, t_j)$. \REFA{We also consider an analogous problem, but given measurements $\{u_j\}_{j=1}^N$ of the velocity field, instead of surface measurements of $h$, such that the cost function}
 \begin{equation}
\mathcal{L}_d = \frac{1}{N_d} \sum_{j\in \Omega_d}(u_j - \hat{u}_j)^2.
\label{loss_general_u}
\end{equation}
\REFA{is minimized}. Below we show how the optimization problem can be solved with Physics-Informed Neural Networks and through the Adjoint State method. \LUCAS{The optimization problem for the 1D cases were solved with both the Adjoint State Method and the Physics-Informed Neural Network method, while the problem for the 2D case was only solved with the latter.}

\subsection{Physics-Informed Neural Networks}

Physics-Informed Neural Networks (PINNs) are neural networks whose training is regularized by the residuals of a system of differential equations of interest \cite{kardianakis,pinn-pathology}, in this case the Shallow Water Eqs. \eqref{SW-u} and \eqref{SW-h}. Under this framework, neural networks, typically, fully connected multilayer perceptrons, are used to parametrize the relevant fields as a function of the coordinates. Here we use two separate networks. One parametrizes $u$ and $h$ and takes $x$ and $t$ as inputs, while the other parametrizes $h_b$ and only takes $x$ as an input. We refer to the output of the networks as $\hat{u}$, $\hat{h}$ and $\hat{h}_b$ and reserve the unhatted variables to the solutions of the SW equations, i.e., our ground-truth data.

\REFA{We first start by describing the 1D case.} Following the setup described in the previous section, the training process
consists of optimizing the set of network parameters
$\theta$ to minimize the cost function 
\begin{equation}
    \mathcal{L}_{\mathrm{PINN}} = \mathcal{L}_d + \lambda_p( \mathcal{L}_u +  \lambda_h \mathcal{L}_h), 
\label{pinn_residuals}
\end{equation}
where $\mathcal{L}_d$ is defined in Eq.~\eqref{loss_general},  with $\{\hat{h}_j\}_{j=1}^N$ being the 
predictions of the PINN at the training points $\Omega_d = \{ x_j , t_j \}_{j=1}^N$. \REFA{Analogously, $\mathcal{L}_d$ is defined in Eq.~\eqref{loss_general_u} if velocity measurements are given instead.} The other two terms of the loss function are

\begin{equation}
    \mathcal{L}_u = \frac{1}{N_p}\sum_{j \in \Omega_p} (\partial_t \hat{u}_j + \hat{u}_j\partial_x\hat{u}_j + g \partial_x\hat{h}_j)^2
    \label{lossphy}
\end{equation}    
which are the residuals of Eq.~\eqref{SW-u} \REFA{evaluated at collocation points $\Omega_p = \{ x_j ,t_j \}_{j=1}^{N_p}$}, and 
\begin{equation}    
\mathcal{L}_h = \frac{1}{N_p}\sum_{j \in \Omega_p} (\partial_t \hat{h}_j + \partial_x( \hat{u}_j (\hat{h}_j-\hat{h}_{bj})))^2,
\label{lossphy2}
\end{equation}
\REFA{which are the residuals of Eq.~\ref{SW-h} evaluated at $\Omega_p$. It is worth noting that the collocation points $\Omega_p$ are not necessarily the same point points as in $\Omega_d$, but they may overlap: we give more details below}.
Both networks share the same loss function, and their training is coupled via $\mathcal{L}_h$. All derivatives are calculated via automatic differentiation over the networks. The terms $\lambda_p$ and $\lambda_h$ are hyperparameters that balance the data and physics parts of the loss functions.
In summary, the estimations $\hat{h}_b$ and $\hat{u}$ are incorporated only in Eqs. \eqref{lossphy} and \eqref{lossphy2}, and optimized to satisfy the Eqs. \eqref{SW-u} and \eqref{SW-h} only, while $\hat{h}$ is optimized to satisfy these equations as well as available data through $\mathcal{L}_d$. \LUCAS{Further details on number of layers, hidden units, learning rates, balance hyperpameters $\lambda_p$ and $\lambda_h$ are given below}.
\REFA{The 2D case is constructed analogously by including the new terms in $\mathcal{L}_u$ and $\mathcal{L}_h$, defining $\mathcal{L}_v$ as the residuals of the other component of Eq.~\eqref{SW-u}.}


\subsection{Adjoint State Method}

\newcommand{\Lagr}{\mathcal{L}}
\newcommand{\Lagradj}{\mathcal{L}_\mathrm{adj}}
\newcommand{\ud}{\tilde{u}^{\dagger}}
\newcommand{\vd}{\tilde{v}^{\dagger}}
\newcommand{\hd}{\tilde{h}^{\dagger}}
\newcommand{\uh}{\tilde{u}}
\newcommand{\vh}{\tilde{v}}
\newcommand{\hh}{\tilde{h}}
\newcommand{\dt}{\partial_t}
\newcommand{\dtau}{\partial_\tau}
\newcommand{\dx}{\partial_x}
\newcommand{\dy}{\partial_y}
\newcommand{\grad}{\partial_x}
\newcommand{\Div}{\partial_x}
\newcommand{\Int}{\int_{0}^{2\pi} \int_{0}^{T}}
\newcommand{\DX}{\,dx\  \,dt}

In this section, we develop the formulation of the Adjoint State Method (ASM) for this problem, as an alternative to PINNs  \cite{minimal-turbulent-channel-flow,adjoint-state-geophysical}. \REFA{We use $\tilde{u}$, $\tilde{h}$ and $\tilde{h}_b$ to denote the outputs of the method. We only show the derivation for the 1D case with height measurements.} The goal of this method is to find a way to calculate $\partial \Lagr_d/\partial \tilde{h}_b$ while enforcing the constraints of satisfying Eqs.~\eqref{SW-u} and~\eqref{SW-h} through the introduction of adjoint state variables, $\ud$ and $\hd$. We assume periodic boundary conditions over $x$ for simplicity and ease of implementation, \REFA{but this is no way fundamental to the derivation of the method.} We take $0$ and $T$ as the minimum and maximum times inside the assimilation window defined by $\Omega_d$. Given this setup, the loss function reads

\begin{equation}
\Lagradj = \Lagr_d + \langle \ud \; , \; \dt \uh + \uh\grad \uh + g \grad \hh \rangle + \langle \hd \; , \; \dt \hh + \Div (\uh (\hh-\hh_b)) \rangle, 
\label{lagrangian}
\end{equation}
where $\langle , \rangle$ is an inner product defined over the whole spatio-temporal domain by

\begin{equation} \langle \mathcal{A}\;,\;\mathcal{B} \rangle = \iint
\!\mathcal{A}\,\mathcal{B}\,
dx\,dt .
\label{scalar-product}
\end{equation}
The optimality conditions are thus given by
\begin{equation}
\frac{\partial\Lagradj}{\partial\hh_b} = 0,
\label{lagr_hb}
\end{equation}
\begin{equation} \frac{\partial\Lagradj}{\partial\uh} = 0 \;,\; \frac{\partial\Lagradj}{\partial\hh} = 0,
\label{lagr_uh}
\end{equation}
\begin{equation}
\frac{\partial\Lagradj}{\partial\ud} = 0 \;,\; \frac{\partial\Lagradj}{\partial\hd} = 0.
\label{lagr_uhd}
\end{equation}
The last two conditions are simply the SW equations, Eqs. \eqref{SW-u} and \eqref{SW-h}. The other conditions can by calculated by first
using the chain rule on Eq.\eqref{lagrangian} to conjugate the inner products, obtaining
\begin{equation}
    \begin{aligned}
    \Lagr_\mathrm{adj} =  \Lagr_d
    &+ \int_{0}^{2\pi} 
    (\uh\ud  + \hh\hd )\bigg|^T_0  \,dx\
    \\
    &- \langle \uh \;,\; \dt \ud +\partial_x(\uh\ud)\rangle
    \\
    &- \langle
    h\;,\; g \Div \ud + \dt \hd \rangle
    - \langle (\hh - \hh_b) \;,\; \uh\grad\hd \rangle,
    \end{aligned}
    \label{lagrangian-adjoint}
\end{equation}
and then differentiating. Condition Eq.~\eqref{lagr_hb} yields

\begin{equation}
\frac{\partial\Lagradj}{\partial\hh_b} = \frac{\partial \Lagr_d}{\partial \hh_b} -
\langle -1 \;,\; \uh \Div \hd \rangle = 0,
\label{df_dhb}
\end{equation}
and thus an expression for the gradient
\begin{equation}
\frac{\partial \Lagr_d}{\partial \hh_b} =
- \int_0^T \uh \partial_x \hd dt ,
\label{df_dhb}
\end{equation}
which is the term of interest, \REFA{while conditions Eq.~\eqref{lagr_uh} yield the equations for the evolution of the adjoint states}

\begin{align}
 \dtau \hd + \uh\grad\hd + g \Div \ud  &= \sum_{j \in \Omega_d} 2 (\hh_j - h_j) ,
 \label{adj_h}
 \\
\dtau \ud + \Div(\uh\ud) + \uh\dx\ud + (\hh - \hh_b)\dx\hd &= 0. 
 \label{adj_u}
\end{align}
It is worth noting that these equations are in reverse time \REFA{$\tau=T-t$, which goes from $T$ to $0$. The initial conditions for the adjoint field can be obtained by noting that
$\Lagr_d$ does not depend explicitly on $\uh(t=T)$ and $\hh(t=T)$, therefore, optimality conditions
\begin{equation}
     \frac{\partial\Lagr_\mathrm{adj}}{\partial \uh(t=T)} = 0 \;,\; \frac{\partial\Lagr_\mathrm{adj}}{\partial \hh(t=T)} = 0
     \label{cond_init_cond}
\end{equation}
only involve the second term in Eq.~\eqref{lagrangian-adjoint}, and yield
\begin{equation} 
\ud(\tau=0) = 0 \;,\;  \hd(\tau=0) = 0,
\label{uadjoint-tau} 
\end{equation}
which are the initial conditions (in reverse time) for the adjoint fields.}


Finally, since initial conditions $u(t=0)=u_0$ and $h(t=0)=h_0$ are unknown, one needs to optimize for these as well by imposing  
\begin{equation}
    \frac{\partial\Lagr_\mathrm{adj}}{\partial \uh_0} = 0 \;,\; \frac{\partial\Lagr_\mathrm{adj}}{\partial \hh_0} = 0.
\label{lagr_uhd}
\end{equation}
Following the same argument used for conditions in Eq. ~\eqref{cond_init_cond}, we now obtain
\begin{equation} 
  \frac{\partial\Lagr_d}{\partial \uh_0}= \ud(x,t=0) \;,\; \frac{\partial\Lagr_d}{\partial \hh_0}= \hd(x,t=0).
\label{df_du0-df_dh0}
\end{equation} 
These expressions, along with Eq. \eqref{df_dhb}, give the direction gradient to be used in the optimization loop, which we describe below.

The optimization problem to find $\hat{h}_b$ (and $\hh_0$ and $\uh_0$) can then be solved by implementing the following algorithm \cite{fractional-step}:

\begin{enumerate}
\item\label{alg1} Solve the SW system in Eqs. \eqref{SW-u} and \eqref{SW-h} using an initial ansatz for $\tilde{h}_b$, $\hh_0$ and $\tilde{u}_0$ from $t=0$ to the final simulation time $t=T$, storing the fields at each time step.

\item \REFA{Solve the adjoint system of equations defined by Eqs. \eqref{adj_u} and \eqref{adj_h} backwards in time, from $\tau=0$ to $\tau=T$, respecting the initial adjoint state conditions at $\tau=0$ and using the fields $\hh$ and $\uh$ obtained in the step above.}

\item Calculate gradients in Eqs. \eqref{df_dhb} and \eqref{df_du0-df_dh0} using $\uh$, $\ud$ and $\hd$, and use them in an optimization algorithm to update the values of $\hh_b$, $\uh_0$ and $\hh_0$, \REFA{for example,
\begin{align*}
\hh_b &\leftarrow \hh_b - \frac{\partial\Lagr_d}{\partial \hh_b},
\\
\uh_0 &\leftarrow \uh_0 - \frac{\partial\Lagr_d}{\partial \uh_0},
\\
\hh_0 &\leftarrow \hh_0 - \frac{\partial\Lagr_d}{\partial \hh_0}.
\label{df_du0-df_dh0}
\end{align*}
}

\item With the updated values, return to step \ref{alg1} of the algorithm, and repeat until convergence. \end{enumerate}

The cost of calculating the gradient required for performing one step in the gradient descent using the adjoint state is equivalent to performing one \textit{forward} integration of the SW equations, plus one \textit{backward} integration of the adjoint state equations. Details on the numerical implementation and the optimization algorithm used are given below.

\section{Numerical experiments}

\subsection{Numerical Simulations and Dataset Generation}

One \REFA{and two} dimensional numerical simulations of Shallow Water flows were performed using a pseudo-spectral scheme to solve Eqs. \eqref{SW-u} and \eqref{SW-h} in a $2\pi$-periodical domain with $1024$ grid point resolution. Numerical stability demands that simulation time step $\Delta t$ satisfy the \textit{CFL} condition, that is:
\begin{equation}
    \Delta t < \frac{\Delta x }{c},
    \label{cfl}
\end{equation}
being $c\sim\sqrt{g\,h_0}$ the characteristic velocity of wave propagation in the linearized case of Eqs. \eqref{SW-u} and \eqref{SW-h} \cite{pedlosky}. To eliminate spurious modes arising from aliasing, the 2/3 rule was applied: all modes satisfying $k > N/3$ were set to zero after performing each pseudo-spectral scheme to compute spatial partial derivatives of a nonlinear term. A second-order Runge-Kutta scheme was implemented for time integration.

\subsubsection{1D numerical simulation}

For \REFA{the one dimensional cases}, we used a Gaussian wave packet as initial conditions
\begin{equation}
    \begin{aligned}        
    h(x,t=0) = h_0 + \eta_0\, e^{- (\frac{x - x_0}{\sigma})^2}\\
    u(x,t=0) = \sqrt{\frac{g}{h_0}} \eta_0 \, e^{- (\frac{x - x_0}{\sigma})^2},
    \end{aligned}
    \label{u_func}
\end{equation}
and a bottom topography given by
\begin{equation}
    h_b(x) = h_{b0}^{(1)}\, \exp\left(- \left(\frac{x - x_0^{(1)}}{\sigma^{(1)}}\right)^2\right) + h_{b0}^{(2)} \exp\left(- \left(\frac{x - x_0^{(2)}}{\sigma^{(2)}}\right)^2\right) + \sum_{j=0}^{60} A_j\, \sin(k_j\,x + \phi_j),
    \label{hb_func}
\end{equation}
where wavenumbers $k_j$ and phases $\phi_j$ were chosen at random from the uniform distributions \(\mathcal{U}_k[40,100]\) and \(\mathcal{U}_{\phi}[0,\pi]\). Each amplitude $A_j$ was chosen at random from values between $1/4$ and $1/2$ the wavelength corresponding to each $k_j$. This profile is plotted in Figure~\ref{hb_croquis}. The values of the all the aforementioned parameters are given in Table \ref{simulation_parameters}.


\begin{table}[t]
\centering
\begin{tabular}{cccccccccc}
$g$ & $h_0$ & $\eta_0$ & $\sigma$ & $\sigma^{(1)}$ & $\sigma^{(2)}$ & $x_0^{(1)}$ & $x_0^{(2)}$ & $h_{b0}^{(1)}$ & $h_{b0}^{(2)}$ \\
\hline
$5$ & $0.2$ & $5\times10^{-5}$ & $\frac{1}{2}$ & $ 0.3$ & $ 0.2$ & $ \frac{\pi}{1.4}$ & $\frac{\pi}{0.8}$ & $0.1$ & $0.05$ \\
\end{tabular}
\caption{Parameters used to generate 1D data.}
\label{simulation_parameters}
\end{table}

When comparing quantities pertaining fields and distances from numerical simulations to the length of the simulated wave, that is $L=2\pi/3$, we obtain relationships that are consistent with the SW hypothesis that $L$ be larger than vertical quantities. For $h$, $h_0/L = 0.095$, and $\eta_0/L=2.40\times10^{-5}$. For $h_b$, amplitudes satisfy $h_{b0}^{(1)}/L=0.048$ and $h_{b0}^{(2)}/L=0.024$.
\REFA{The setup presented here resembles that of a wave approaching the shore (but before the wave breaking) or of a tsunami traveling through the wider ocean. In order to give more physical context to these experiments, we now compare the characteristic scales of our simulations with that of tsunamis.}
The width of the simulated wave can be mapped to the real physical scale of such a flow by defining that \(\sigma \sim 10000m\) \cite{tsunamis}. Since the \(\sigma\) used yields a wave length approximately $L=2\pi/3$, the mapping implies that the span of the $2\pi$-periodic domain is \(N\,\Delta x \sim 30000m\). This maps distance values in the simulation to physical units. Thus, the distance between grid points $\Delta x$ is mapped to $29.3m$. The initial wave height is mapped to \(\eta_0 \sim 1m\). The \(h_b\) main amplitudes are mapped to \(h_{b0}^{(1)} \sim 2094m\) and \(h_{b0}^{(2)} \sim 1047m\). The two main structures present in $h_b$ will have their characteristic horizontal dimensions, $\sigma^{(1)}$ and $\sigma^{(2)}$, mapped to $6000m$ and $4000m$, respectively. The mean height from a flat bottom represents \(h_0 \sim 4200m\). The smaller structures present \(h_b\) are mapped to lengths \(l \in [300m,\,770m]\).
Total simulation time is $T = 5$ with a time step $\Delta t=1\times10^{-4}$. Time is normalized by $t_L=L/c$, which is the time for the propagating wave to travel one wave length. Thus, $T/t_L = 2.39$: just short of letting the wave propagate through the entire domain.





\begin{table}[t]
\centering
\begin{tabular}{c|ccccc}
$n_x$     & 1 & 16 & 64 & 128 \\
\hline
$\delta_x$ & $3/1024$ & $3/64$ & $3/16$ & $3/8$ \\
\hline
$l_x$ [m] & 60 & 470 & 1900 & 3800 \\
\end{tabular}
\caption{Separation between measurements, in number of grid points $n_x$, normalized distances $\delta_x$, and approximate physical distances $l_x$}
\label{spacings}
\end{table}

\begin{table}[h]
\centering
\begin{tabular}{c|ccccc}
$\epsilon$ & $1\times 10^{-7}$ & $5\times 10^{-7}$ & $1\times 10^{-6}$ & $5\times 10^{-6}$ \\
\hline
$\epsilon/\eta_0$ & $2\times 10^{-3}$ & $1\times 10^{-2}$ & $2\times 10^{-2}$ & $1\times 10^{-1}$ \\
\end{tabular}
\caption{Values of standard deviation $\epsilon$ for measurement noise, and relation to initial wave amplitude, $\epsilon/\eta_0$.}
\label{noise}
\end{table}

We perform experiments with different levels of sparsity in our simulated measurements. Table~\ref{spacings} shows the different values used in terms of grid points, $n_x$, normalized distance $\delta_x = n_x \Delta x /L$ and approximate physical distance (according to the values presented above), $l_x$.
Additionally, for a fixed $\delta_x= 3/64$, noise is added to surface height data to test the robustness of each method. We use noise sampled from a Gaussian distribution $\mathcal{N}(0,\epsilon)$, with various standard deviations, $\epsilon$ that yield increasing amplitudes for the added noise. Table \ref{noise} shows the values of $\epsilon$, along with their relation to the initial wave amplitude $\eta_0$.

\subsubsection{\REFA{2D numerical simulations}}


\REFA{The 2D case was built as follows. The initial condition consists of a Gaussian wave packet with translational symmetry in the $y$ direction, propagating in the $x$ direction of the domain:}

\begin{equation}
    \begin{aligned}        
    h(x,y,t=0) = h_0\prime + \eta_0\prime\, e^{- (\frac{x - x_0\prime}{\sigma\prime})^2}\\
    u(x,y,t=0) = \sqrt{\frac{g\prime}{h_0\prime}} \eta_0\prime \, e^{- (\frac{x - x_0\prime}{\sigma\prime})^2},
    \end{aligned}
    \label{u_func}
\end{equation}
\REFA{for all $y$ in the $2\pi$-periodic domain. The bottom topography was taken as} 

\begin{equation}
    h_b(x,y) = h_{b0} \, \cos(2\pi \, 5x) \, \cos(2\pi \, 5y).
\end{equation}

\REFA{The values of the all the aforementioned parameters for 2D numerical simulations are given in Table \ref{2Dsimulation_parameters}. A diagram for this setup is shown in figure~\ref{dg2d_diagram}.}
\REFA{Horizontal distances are normalized by the wave packet characteristic $L=2\pi/4$, field $h$ amplitudes are normalized by initial wave amplitude $\eta_0 = 0.05$ and $h_b$ amplitudes are normalized by $h_0$, analogously as with the 1D case. Total simulation time is $T = 6$ with frames saved every time step $\Delta t=2.5\times10^{-2}$. Time is normalized by $t_L=L/c$, which is the time for the propagating wave to travel one wave length. Thus, $T/t_L = 3.8$, just short of letting the wave propagate through the entire domain.}


\begin{table}[t]
\centering
\begin{tabular}{cccccc}
$g\prime$ & $h_0\prime$ & $\eta_0\prime$ & $\sigma\prime$ & $x_0\prime$ & $h_{b0}$ \\
\hline
$1$ & $1$ & $5\times10^{-2}$ & $\frac{\pi}{8}$ & $\frac{\pi}{2}$ & $0.1$  \\
\end{tabular}
\caption{Parameters used to generate 2D data.}
\label{2Dsimulation_parameters}
\end{table}

\subsection{PINN parameters}\label{PINN parameters}

\begin{figure}[t]
    \centering
         \includegraphics[width=0.8\textwidth]{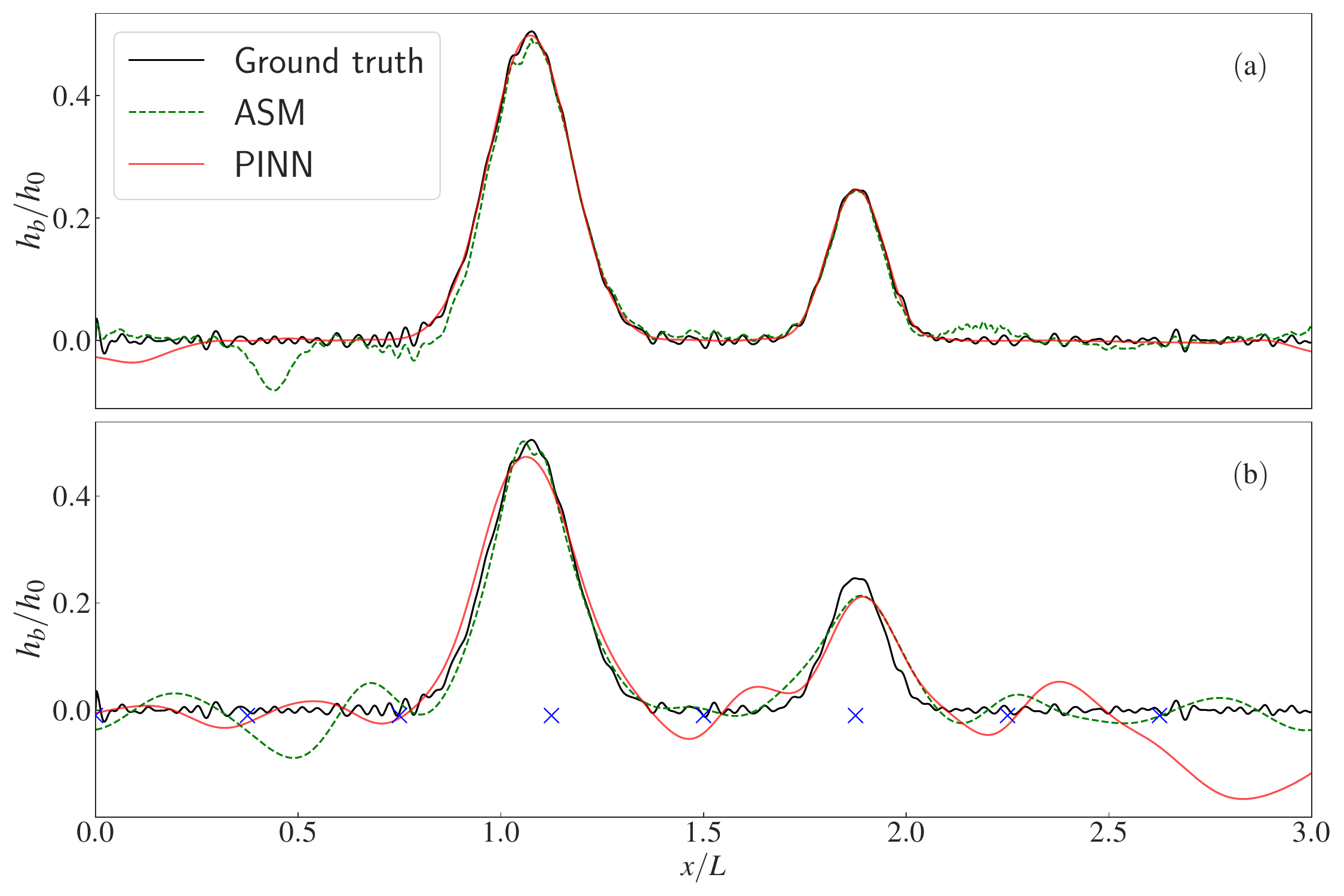}
        \caption{Reconstructed bottom topographies $h_b$ for the cases with distance between measurements of
        (a) $\delta_x=3/1024$ and
        (b) $\delta_x=3/8$.
        The true $h_b$ is marked with black solid line, the green dashed line denotes the ASM results $\tilde{h}_b$ and the red solid line denote PINN results $\hat{h}_b$.}
    \label{hb_no_noise}
\end{figure}

The networks were trained using Adam with a mini-batch size of $m_b=2000$, and a learning rate of $\mu=1\times10^{-6}$ for the 1D flow assimilation, \REFA{and $m_b=10000$ and $\mu=1\times10^{-7}$ for the 2D case}.  Each network was trained during $5990$ epochs for the 1D flow, \REFA{and for $260$ epochs for the 2D flow}. The activation function used was \textit{Siren} \cite{siren} in all cases. 
\LUCAS{Different combinations of the number of layers $l$ and hidden units $w$ for the network that parametrizes $u$ and $h$ are indicated in table \ref{pinn_parameters} for various experiments of different \(\delta_x\) for the 1D flow. Different combinations of layers $l_{inv}$ and hidden units $w_{inv}$ for the network that parametrizes $h_b$ are also indicated in table \ref{pinn_parameters}.} For the sparsest case ($\delta_x = 3/8$) an ensemble of PINNs with different combinations of layers and hidden units were used. Ensembles of PINNs were also used for the cases with added noise to measurements. The parameters used for the networks in these cases are indicated in table \ref{pinn_noise_parameters}. In all other cases, that is $\delta_x =\{3/1024, 3/64/, 3/16\}$, only one PINN is used, per case. The results shown represent the mean of the ensemble.

\REFA{For 2D flow assimilation, only dense $\delta_x$ experiments were performed. A random sampling of collocation and data points was taken at each time frame of the simulation for network training, involving approximately $50\%$ of the domain at each time frame. A different combination of network parameters  $l$, $w$, $l_{inv}$ and $w_{inv}$ were used for these flows as well. These are indicated in table \ref{2Dpinn_parameters}.}

\begin{table}[t]
\centering
\begin{tabular}{c|cccc}
$\delta_x$ & $3/1024$ & $3/64$ & $3/16$ & $3/8$ \\
\hline
$l$ & 4 & 4 & 4 & [4,6] \\
\hline
$l_{inv}$ & 4 & 4 & 4 & 4 \\
\hline
$w$ & 100 & 100 & 100 & [100,150,200] \\
\hline
$w_{inv}$ & 100 & 100 & 100 & 100 \\
\hline
\end{tabular}
\caption{PINN parameters for 1D flow assimilation for various distance $\delta_x$. The brackets denote values used in different ensemble members for the sparsest case.}
\label{pinn_parameters}
\end{table}

\begin{table}[t]
\centering
\begin{tabular}{c|cccc}
$\epsilon$ & $1\times 10^{-7}$ & $5\times 10^{-7}$ & $1\times 10^{-6}$ & $5\times 10^{-6}$ \\
\hline
$l$ & [5,6] & [4,5,6] & [5,6] & [5,6] \\
\hline
$l_{inv}$ & 4 & 4 & 4 & 4 \\
\hline
$w$ & [100,150,200] & [100,150,200] & [100,150,200] & [100,150,200] \\
\hline
$w_{inv}$ & 100 & 100 & 100 & 100 \\
\hline
\end{tabular}
\caption{PINN parameters for 1D flow assimilation for different values of noise amplitude $\epsilon$ added to data. The brackets denote values used in different ensemble members for each case.}
\label{pinn_noise_parameters}
\end{table}

\begin{table}[t]
\centering
\begin{tabular}{c|c}
$l$ & 6 \\
\hline
$l_{inv}$ & 4 \\
\hline
$w$ & 200 \\
\hline
$w_{inv}$ & 100 \\
\hline
\end{tabular}
\caption{PINN parameters for 2D flow assimilation.}
\label{2Dpinn_parameters}
\end{table}


Following \cite{pinn-pathology}, the balance hyperparameters $\lambda_p$ and $\lambda_h$
were updated \REFA{at each epoch} by weighing the gradients of
the loss terms in Eq. \eqref{pinn_residuals} against each other: 
\begin{equation}
            \lambda_p^{i+1} = (1-\alpha)\lambda_p^{i} + \alpha\frac{\langle \nabla_{\theta}L_d^i\rangle }{\langle\nabla_{\theta}L_u^i + \lambda_h^{i+1}\nabla_{\theta}L_h^i\rangle},
            \label{lambdap}
\end{equation}
where $\alpha$ is a \REFA{new} hyperparameter, and where 
\begin{equation}
            \lambda_h^{i+1} = (1-\alpha)\lambda_h^{i} + \alpha\frac{\langle \nabla_{\theta}L_u^i\rangle}{\langle\nabla_{\theta}L_h^i\rangle},
            \label{lambdah}            
\end{equation}
so that the total gradient of Eq. \eqref{pinn_residuals} at epoch $i+1$ yielded
\begin{equation}
    \nabla_{\theta} \mathcal{L}^{i+1}= \nabla_{\theta} L_d^{i+1} + \lambda_p^{i+1}(\nabla_{\theta} L_u^{i+1} + \lambda_h^{i+1}\nabla_{\theta} L_h^{i+1}).
    \label{pinn_residual_gradient}
\end{equation}
\REFA{Following the original reference \cite{pinn-pathology}, we set $\alpha=0.1$.} \REFA{The same procedure was extended to the 2D case in a straightforward fashion.}

\begin{figure}[t]
    \centering
         \includegraphics[width=0.8\textwidth]{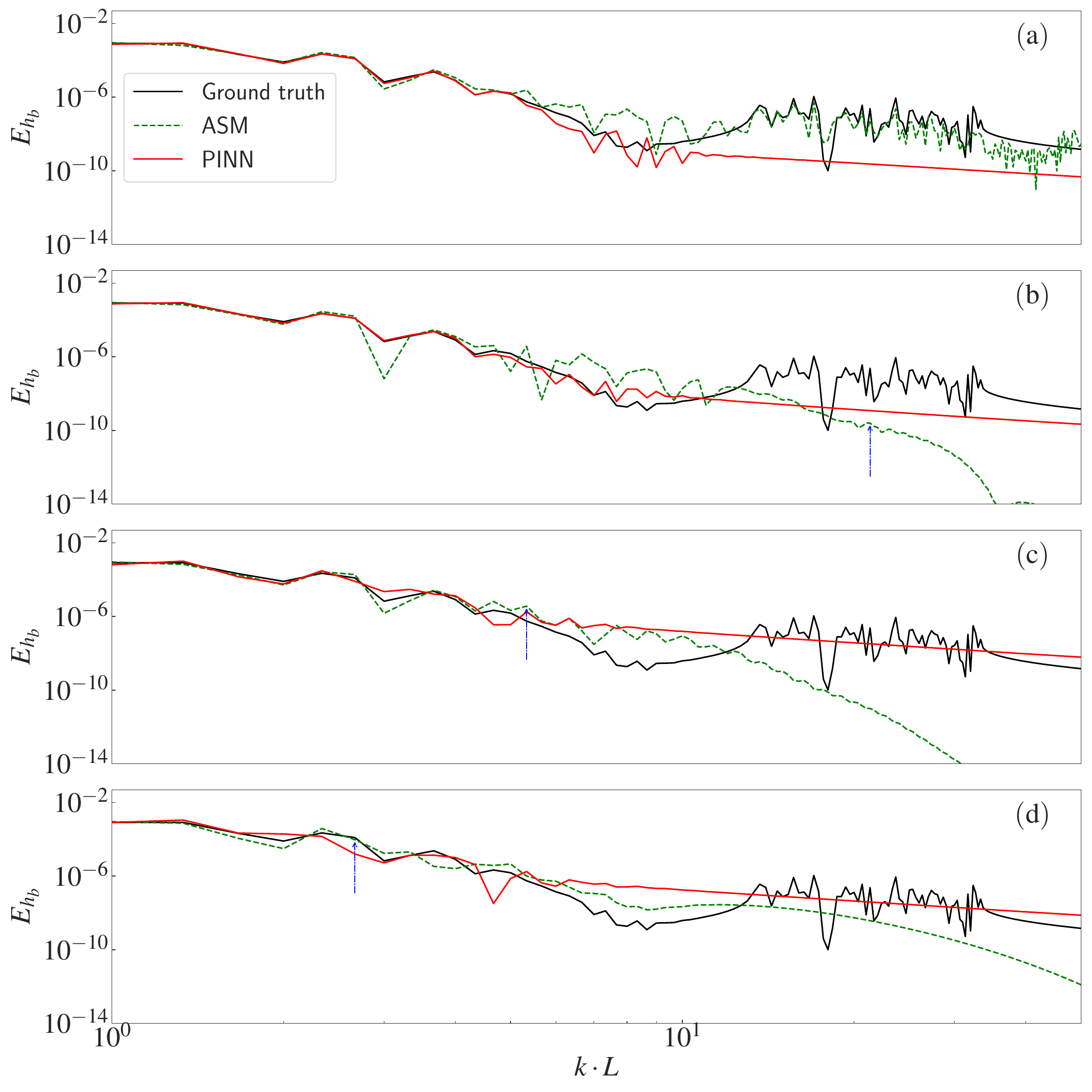}
        \caption{\REFA{Fourier spectra $E_{h_b}$
        of the ground truth (black solid line), 
        the ASM reconstruction (green dashed line),
        and the PINN reconstruction (red solid line).
        Each panel corresponds to cases with different separation between measurements: (a) $\delta_x = 3/1024$, (b) $\delta_x = 3/64$, (c) $\delta_x = 3/16$ and (d) $\delta_x = 3/8$. The blue arrow indicates the wavelength corresponding to separation $\delta_x$ in each case.} }
    \label{ffthb_nx-multiplot}
\end{figure}

\subsection{Adjoint State Method parameters}

The adjoint state equations were solved using the same pseudo-spectral method as for the SW equations described above. The only caveat being that the forcing term was projected using Fourier interpolation so as to avoid discontinuities. \REFA{It is important to mention that the method can be implemented with other numerical schemes and other boundary conditions.}
The L-BFGS algorithm was used for the optimization steps. The gradient descent step was $10^{-1}$ for $h_b$ field optimization, and $10^{-2}$ for initial fields $u(t=0)$ and $h(t=0)$.

The ansatz for the initial height $\hh(t=0)$ is built by interpolating the sparse measurements by fitting them to a Gaussian curve. This is done by optimizing the parameters $A,B,x_0,\sigma$ in 
\begin{equation*}
    B + A \, e^{-(\frac{x-x_0}{\sigma})^2}
\end{equation*}
to the measurements, using a least squares approach. We generate an ansatz for the initial velocity field by taking $\uh_0 = \sqrt{g/h_0} \tilde{\eta}_0$.

\begin{figure}[t]
    \centering
\includegraphics[width=0.8\textwidth]{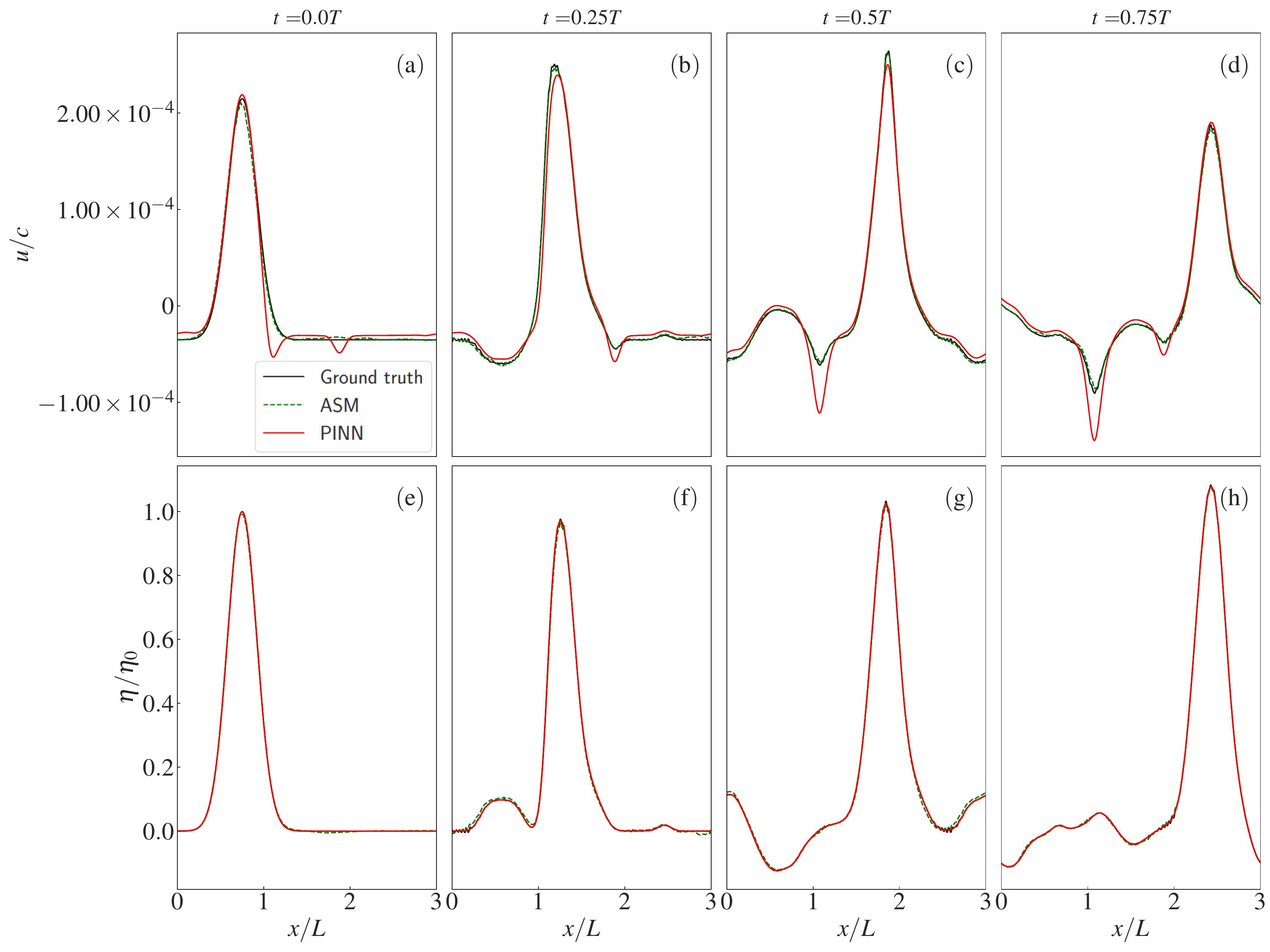}
        \caption{Dense measurement case, $\delta=3/1024$. (a-d): True and reconstructed velocity fields $u/c$ at different times. (e-h): True and reconstructed surface fluctuations $\eta/\eta_0$. Same legends as figure~\ref{hb_no_noise}.}
    \label{u_h_no_noise_nx1}
\end{figure}

\begin{figure}[t]
    \centering
\includegraphics[width=0.8\textwidth]{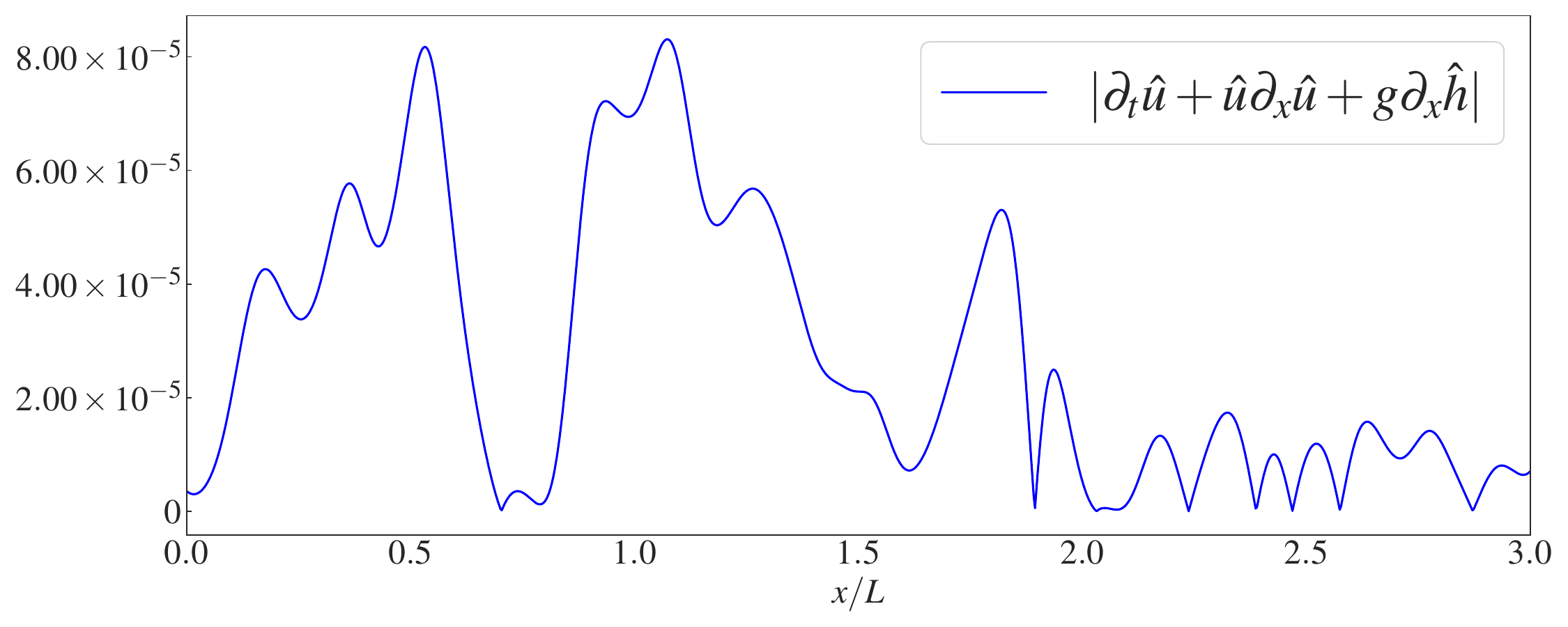}
        \caption{\REFA{Absolute value of the SW momentum equation evaluated on PINN predicted fields at time $0.5\,T$, for dense measurement case $\delta=3/1024$.}}
    \label{residuals1}
\end{figure}

\subsection{Error Metrics}



\REFA{We define the global errors in field reconstruction for the 1D cases as follows. For the topography reconstruction
\begin{equation} 
\mathcal{E}_{h_b} = \int_0^{2\pi} \frac{\sqrt{(h_b(x) - h_b(x)^\prime)^2}}{\max\limits_{x}[h_b(x)]} \,dx,
\label{errorhb}
\end{equation}
and the global errors in the velocity reconstruction as
\begin{equation}
\mathcal{E}_{u} = \int_0^{2\pi}\int_0^{T}  \frac{\sqrt{(u(x,t) - u(x,t)^\prime)^2}}{\max\limits_{x,t}[u(x,t)]}\,dt\,dx.
\label{erroru}
\end{equation}
The primed variables indicate the corresponding reconstruction given by either the PINN or ASM. In the text, we explicitly clarify for which method we are calculating errors.}


\REFA{In order to perform scale-by-scale comparisons, we also study the Fourier spectra of $h_b$, defined as}
\begin{equation}    
E_{h_b} = \norm{\mathcal{F}[h_b/L]}^2
\label{Ehb}\end{equation}
\REFA{where $\mathcal{F}[f]$ is the Fourier transform of field $f$.}

\begin{figure}[h!]
    \centering
 \noindent\includegraphics[width=0.8\textwidth]{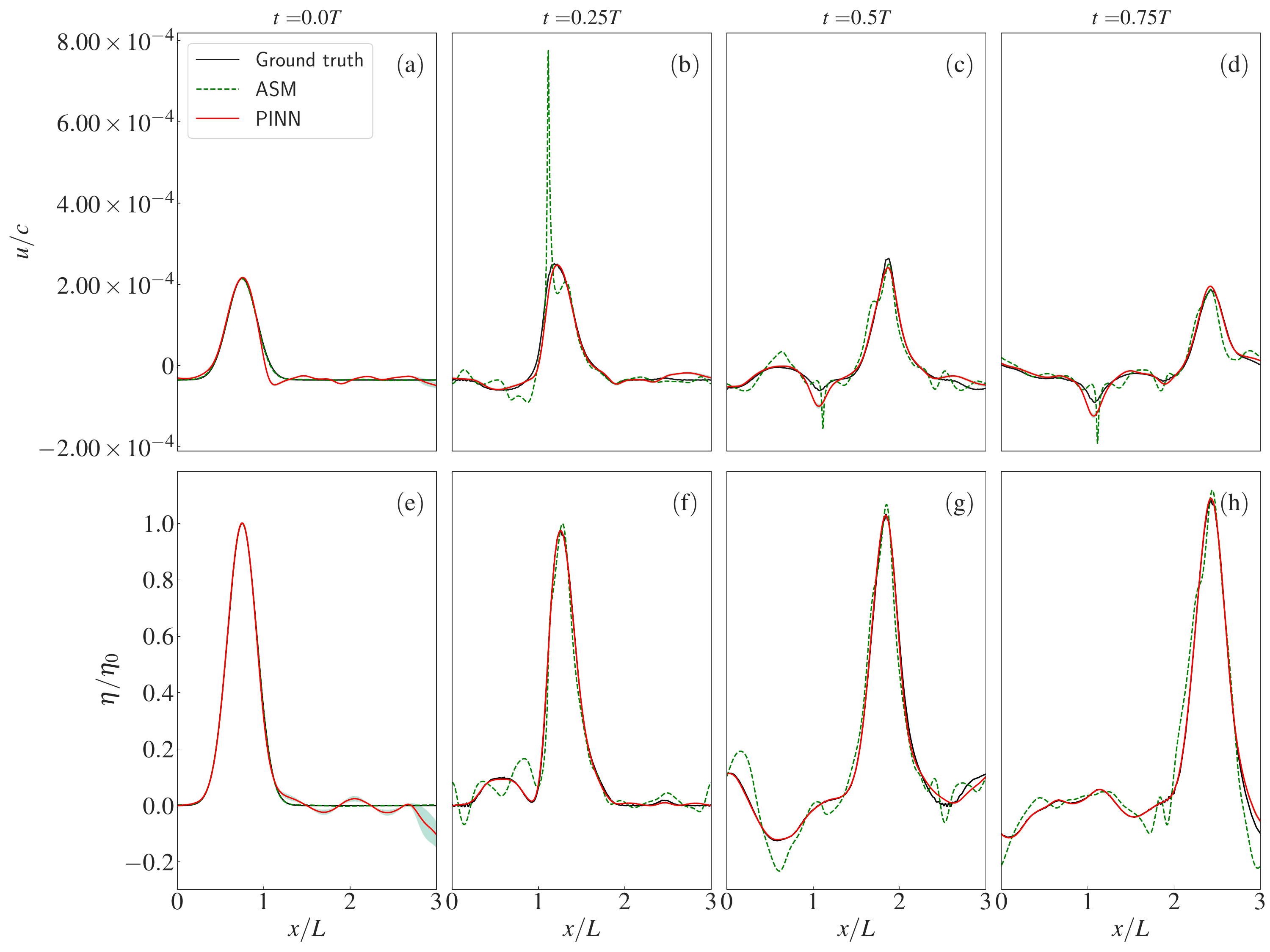}
         \label{u_no_noise_nx128}
        \caption{Sparse measurement case, $\delta=3/8$. (a-d): true and reconstructed velocity fields $u/c$ at different times. (e-h): true and reconstructed surface fluctuations $\eta/\eta_0$. Same legends as figure~\ref{hb_no_noise}. }
    \label{u_h_no_noise_nx128}
\end{figure}

\begin{figure}[t]
    \centering
         \includegraphics[width=0.8\textwidth]{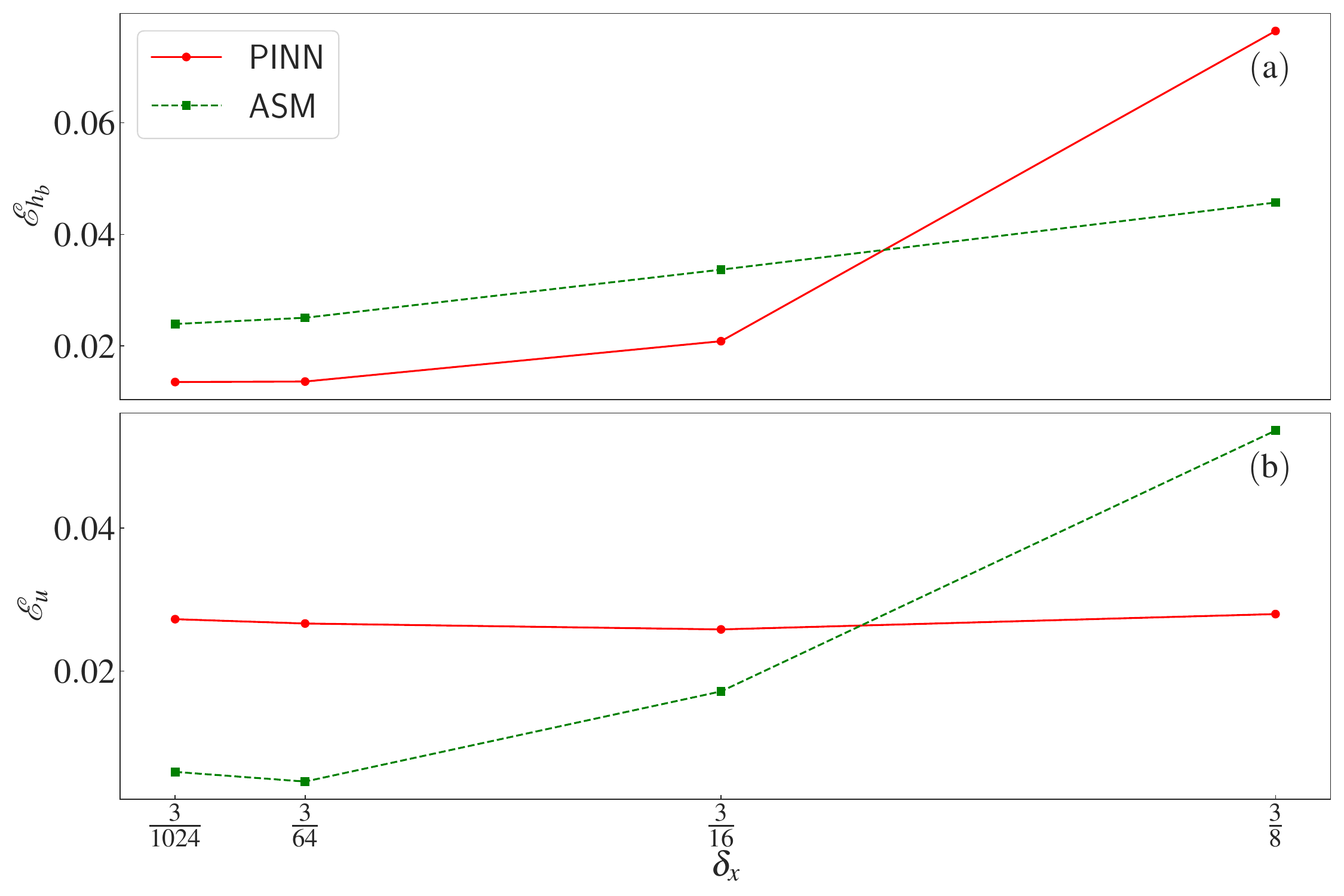}
        \caption{Errors as function of the measurement distance $\delta_x$. (a) Global errors in the topography reconstruction $\mathcal{E}_{{h}_b}$, and (b) global error in the velocity reconstruction $\mathcal{E}_{{u}}$. The errors for the ASM are marked with the green dashed line, while the errors for the PINN are marked with the red solid line.}
    \label{error_sweep-nx}
\end{figure}

\section{Results}
\label{sec:results}

\begin{figure}[t]
    \centering
         \includegraphics[width=0.8\textwidth]{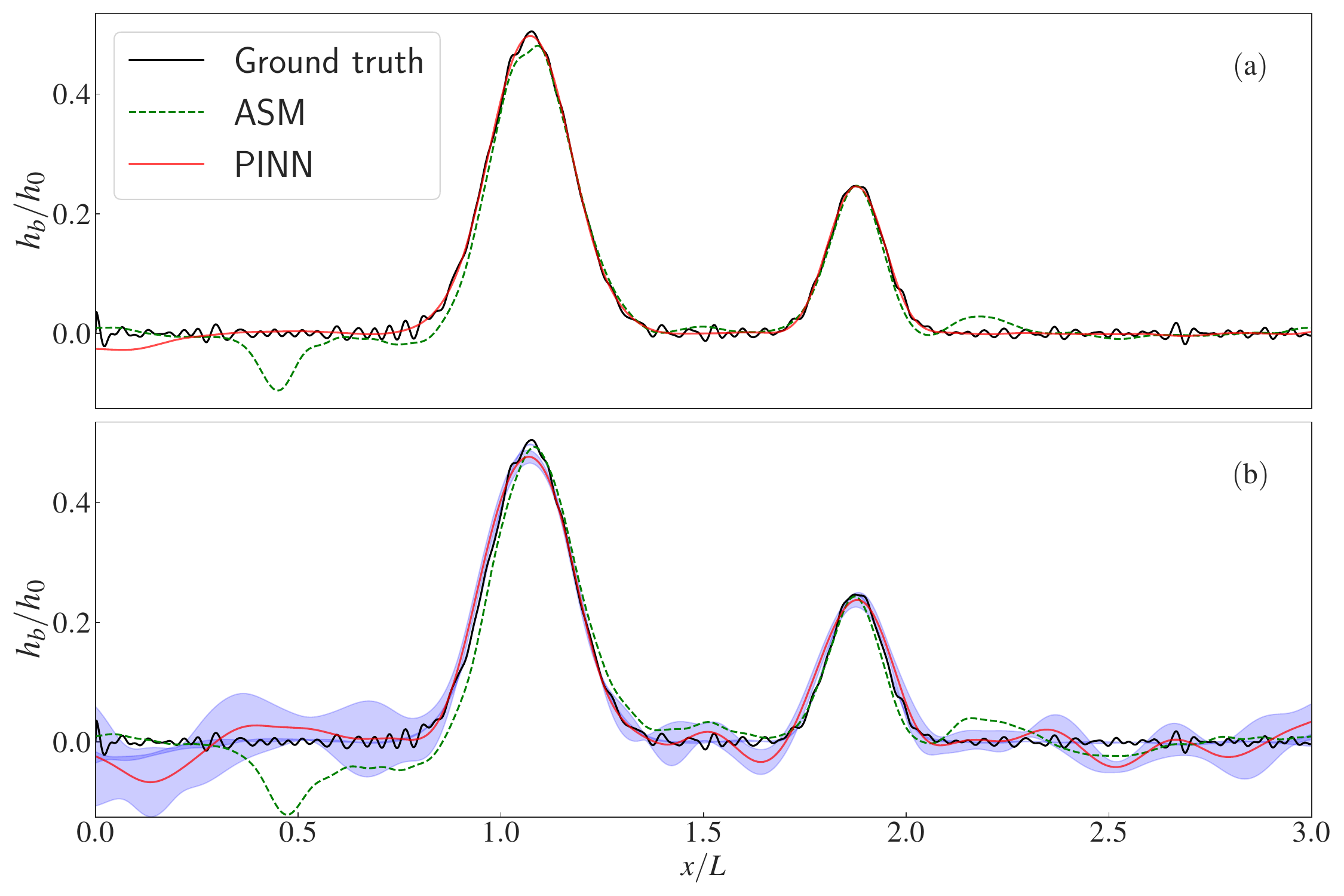}
        \caption{Reconstructed bottom topographies $h_b$ for the cases with noise amplitudes (a) $\epsilon/\eta_0 = 2\times10^{-3}$ and (b) $\epsilon/\eta_0 = 1\times10^{-1}$.
        The true $h_b$ is marked with black solid line, the green dashed line denotes the ASM results $\tilde{h}_b$ and the red solid line denote PINN results $\hat{h}_b$. The blue shade represents the standard deviation between different PINN realizations, for the same data points.}
    \label{hb_noise-multiplot}
\end{figure}

\begin{figure}[h]
    \centering
         \includegraphics[width=0.8\textwidth]{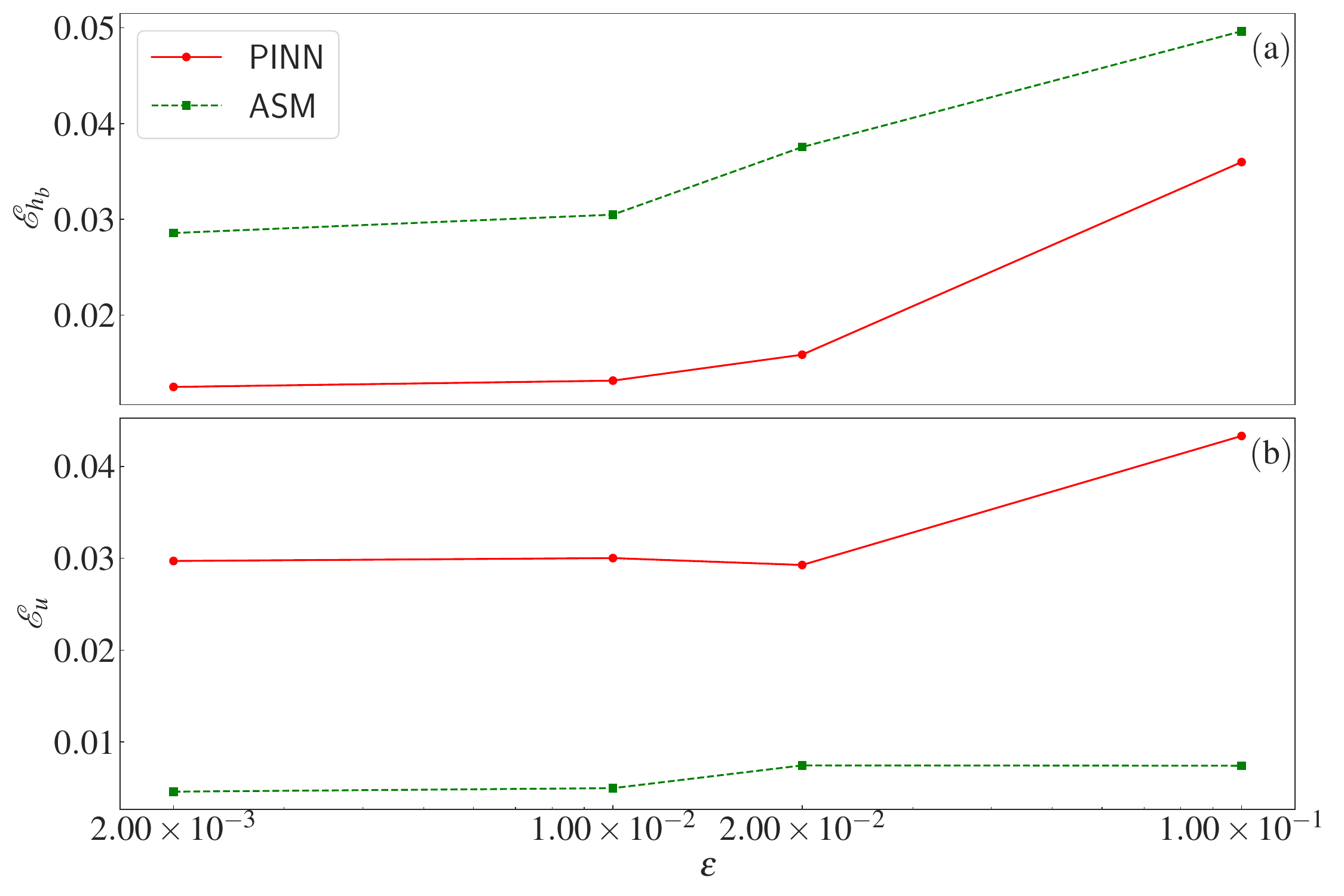}
        \caption{Errors as function of the added noise $\epsilon/\eta_0$. (a) Global errors in the topography reconstruction $\mathcal{E}_{{h}_b}$, and (b) global errors in the velocity reconstruction $\mathcal{E}_{{u}}$. The errors for the ASM are marked with the green dashed line, while the errors for the PINN are marked with the red solid line.}
    \label{error_sweep-noise}
\end{figure}

\subsection{Sparse measurements}
\label{sec:noiseless results}

We begin by showing the results for PINNs and ASM for the case without any added noise, $\epsilon=0$, and for different distances between measurements $\delta_x$.
Figure \ref{hb_no_noise} shows predictions from PINNs $\hat{h}_b$ (red solid line) and from the ASM  $\tilde{h}_b$ (green dashed line), plotted against the true $h_b$ field (black solid line) for  (a) $\delta_x=3/1024$ and (b) $\delta_x=3/8$. Both methods can reconstruct the overall shape of the bottom topography with good accuracy, even in the sparser case. \REFA{The PINN produces, overall, smoother results and has a tendency to misrepresent the boundaries, as they are less data dense. On the other hand, the ASM is able to capture some of the finer details present in the topography. These behaviors are expected, as neural networks usually have difficulties learning high-frequency structures, a problem known as spectral bias \cite{tancik_Fourier_2020}, while the pseudospectral-based ASM does not.}

\REFA{Before we continue analyzing the results, we give details on the computational cost of each method.} The ASM calculations took between 2 and 4 hours, the longest being the less sparse cases, where smaller scales could be resolved. Training the PINNs took from 10 to 24 hours to converge. \REFA{In the 1D case, both methods were solved using the same CPU}. The PINN ensemble shown in panel (b) of figure \ref{hb_no_noise} is made out of 5 $\hat{h}_b$ predictions. Different network architectures and random initializations of the network parameters lead to different results after optimization. The mean of all PINN realizations for a single case is considered.

\REFA{
In order to perform a scale-by-scale comparison, we show in figure~\ref{ffthb_nx-multiplot} the Fourier spectra $E_{h_b}$ of the ground truth, the ASM reconstruction, and the PINN reconstruction for four different measurement distances $\delta_x$. Both methods perform well in reconstructing the largest scales for every measurement distance, although results do deteriorate as $\delta_x$ is increased. Both methods also present errors around $kL\approx 1$, with the PINN obtaining better results. Only the ASM in the densest case is able to reconstruct the smaller scales of the system. \LUCAS{As mentioned above, neural networks can struggle with high-frequency data. While steps were taken to alleviate this (see discussion on activation functions in \cite{siren}), the problem persists.} Further training may produce better results but at a large computational cost \cite{anagnostopoulos_Learning_2024}.}

\begin{figure}[h]
    \centering
         \includegraphics[width=0.8\textwidth]{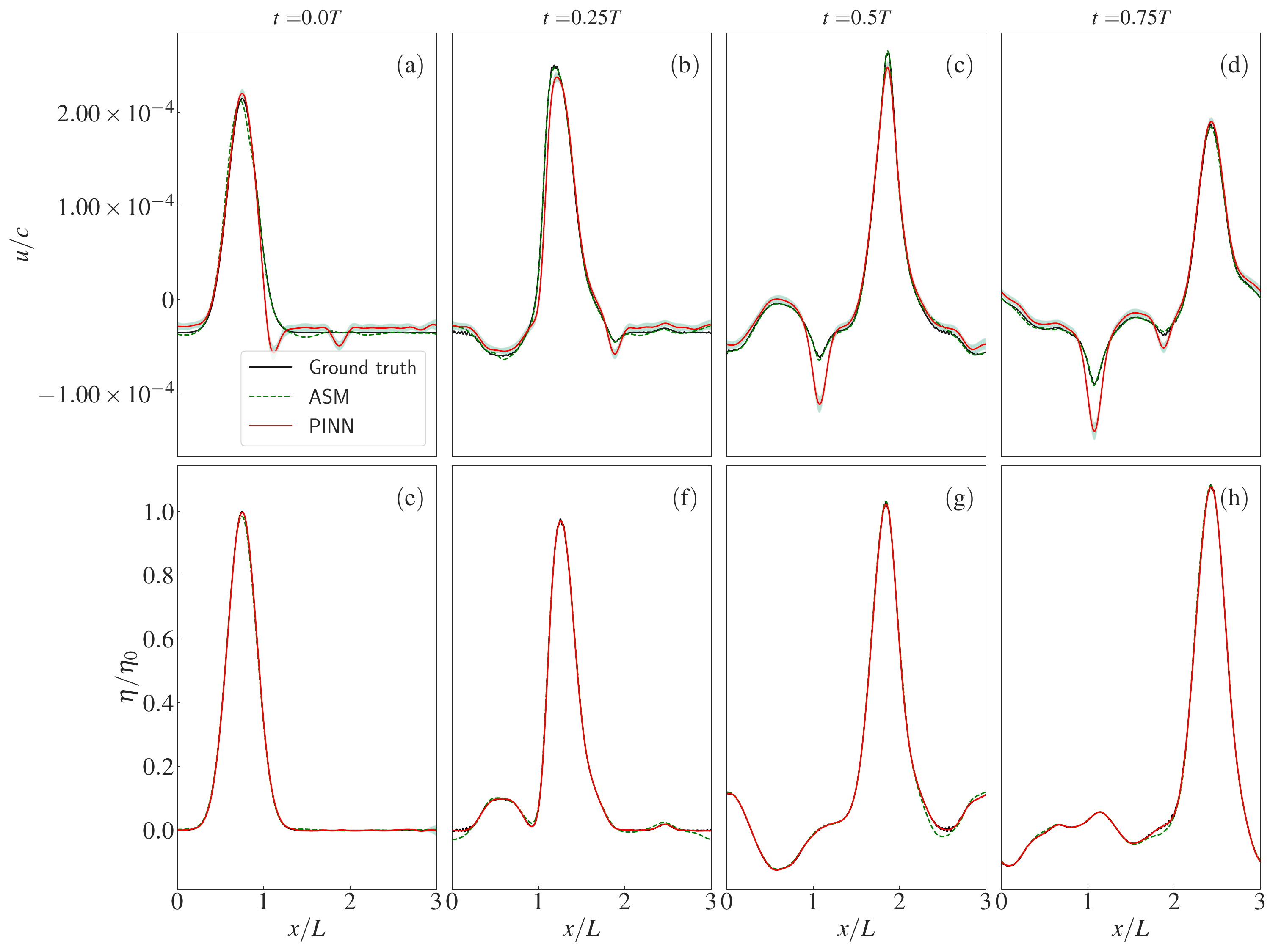}
        \caption{Case with noise amplitude $\epsilon/\eta_0 = 0.02$ (a-d): True and reconstructed velocity fields $u/c$ at different times. (e-h): True and reconstructed surface fluctuations $\eta/\eta_0$. Same legends as figure~\ref{hb_no_noise}.}
    \label{uh_noise}
\end{figure}

Figure \ref{u_h_no_noise_nx1} shows predictions for fields $u$ and $\eta$, for the case with $\delta_x = 3/1024$. It is noticeable that, while both methods are mostly successful in reconstructing the velocity field, the ASM is considerably better, \REFA{as the PINN produces small artifacts near the location of the two peaks.} As for the surface height, both methods are trivially successful in its reconstruction, since both have access to surface height data, in this case, with full spatial resolution.
For the sparser case of $\delta_x=3/8$, shown in figure \ref{u_h_no_noise_nx128}, the ASM presents much larger fluctuations than the PINN, \REFA{which thanks to its difficulty in capturing high-frequency structures produces smoother and more regular fields}. For example, in panel (b) where the wave is traveling over the largest gradient for $h_b$, there appear to be discontinuities in $\tilde{u}$ while $\hat{u}$ remains quite smooth. Similar behavior is observed when reconstructing $\eta$, as seen in panels (e) to (h). \REFA{It is worth noting that the blue shaded region is only appreciable in panel (e) due to the relatively small scale of the PINN ensemble standard deviation in the other panels.}

\REFA{In order to check that the deviations in the velocity prediction obtained by the PINN are in fact non-physical artifacts, we calculated the residuals of the equations of motion. In figure~\ref{residuals1} we show the residuals of the SW momentum equation Eq. \eqref{SW-u} as a function of $x$ evaluated at time $t=0.5T$. As expected the residuals are higher at the beginning of the domain where the wave has already passed, and noticeably, their values around $x/L=1$ are also elevated. Fixing this issue, either by trainig further or by using residual-based sampling \cite{wu_Comprehensive_2022}, should improve the performance of the PINN.}


Finally, figure \ref{error_sweep-nx} shows global errors (a) in the topography reconstruction $\mathcal{E}_{h_b}$ and (b) in the velocity reconstruction $\mathcal{E}_u$, for each sparsity in data points considered. In both methods the error in reconstructing the bottom topography increases as the separation between measurements increases, as expected, with PINNs outperforming ASM when measurements are dense and ASM outperforming PINNs when they are sparse. \REFA{Although it should be noted that the PINN results at the lower sparsity are dominated by the errors in edges of the domain, as can be seen in figure~\ref{hb_no_noise}}. These results are reversed when looking at the errors of the reconstruction of the velocity fields in panel (b) where ASM achieves better results than PINN when measurements are dense. Interestingly, the error in the PINN reconstruction is stable with respect to measurement distance in this case.

\subsection{Noisy measurements}
\label{sec:noise results}

This subsection discusses the results of PINNs and ASM when assimilating data with different amplitudes of added noise. In all cases presented here the distance between measurements is $\delta_x= 3/64$. Figure~\ref{hb_noise-multiplot} shows the bottom topography reconstructed by the PINN and by the ASM for the cases with (a) $\epsilon/\eta_0 = 0.002$ and (b) $\epsilon/\eta_0 = 0.1$. Both methods are able to reconstruct the main structures correctly, although PINNs achieves a more accurate result. It is interesting that in the noisier case the different PINN outcomes have much greater variance than in Figure~\ref{hb_no_noise}.

\begin{figure}[t]
    \centering
         \includegraphics[width=0.7\textwidth]{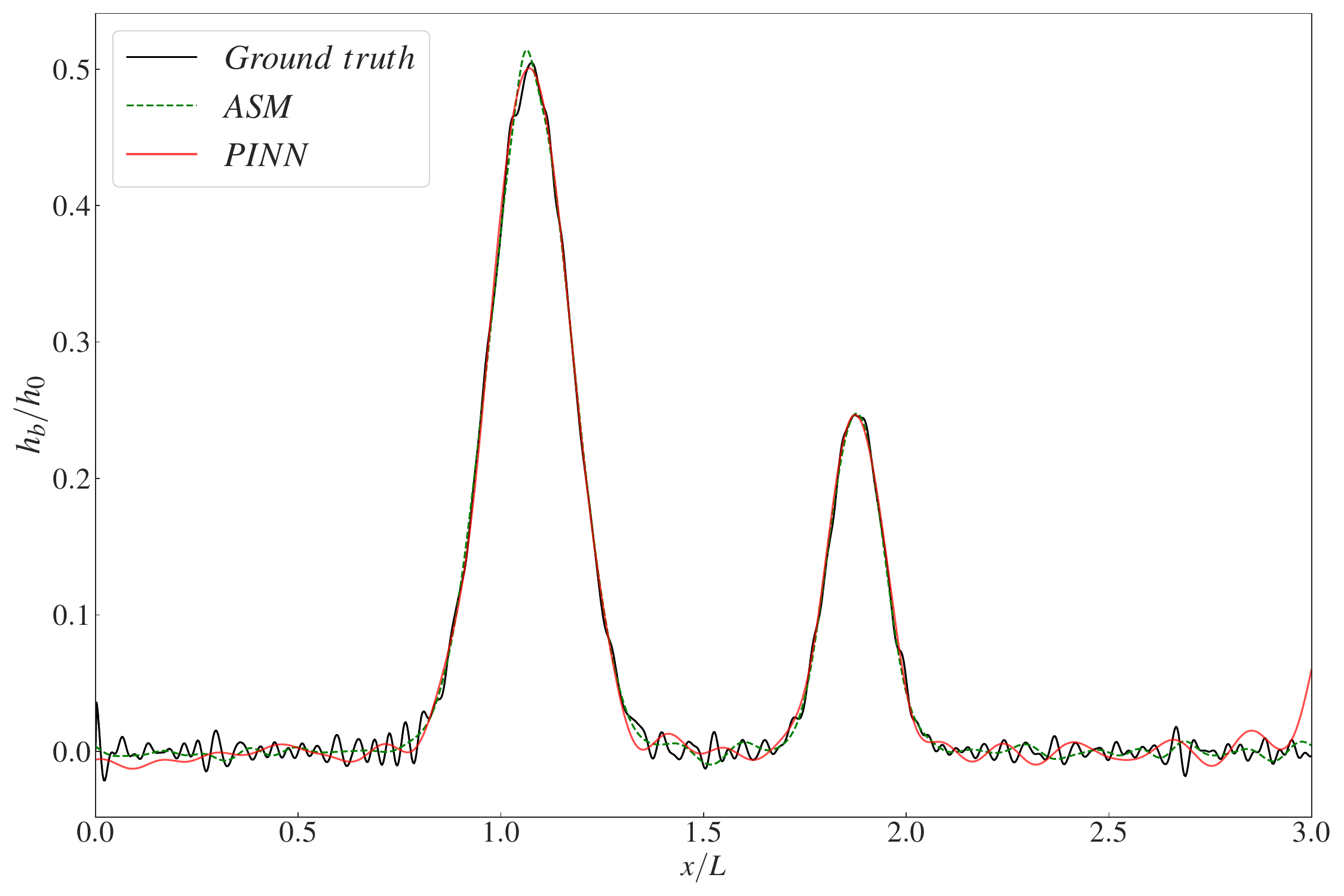}
        \caption{Reconstructed bottom topographies $h_b$ using velocity measurements instead of height measurements. The true $h_b$ is marked with black solid line, the green dashed line denotes the ASM results $\tilde{h}_b$ and the red solid line denote PINN results $\hat{h}_b$.}
    \label{udata_hb}
\end{figure}

\begin{figure}[h]
    \centering
         \includegraphics[width=0.8\textwidth]{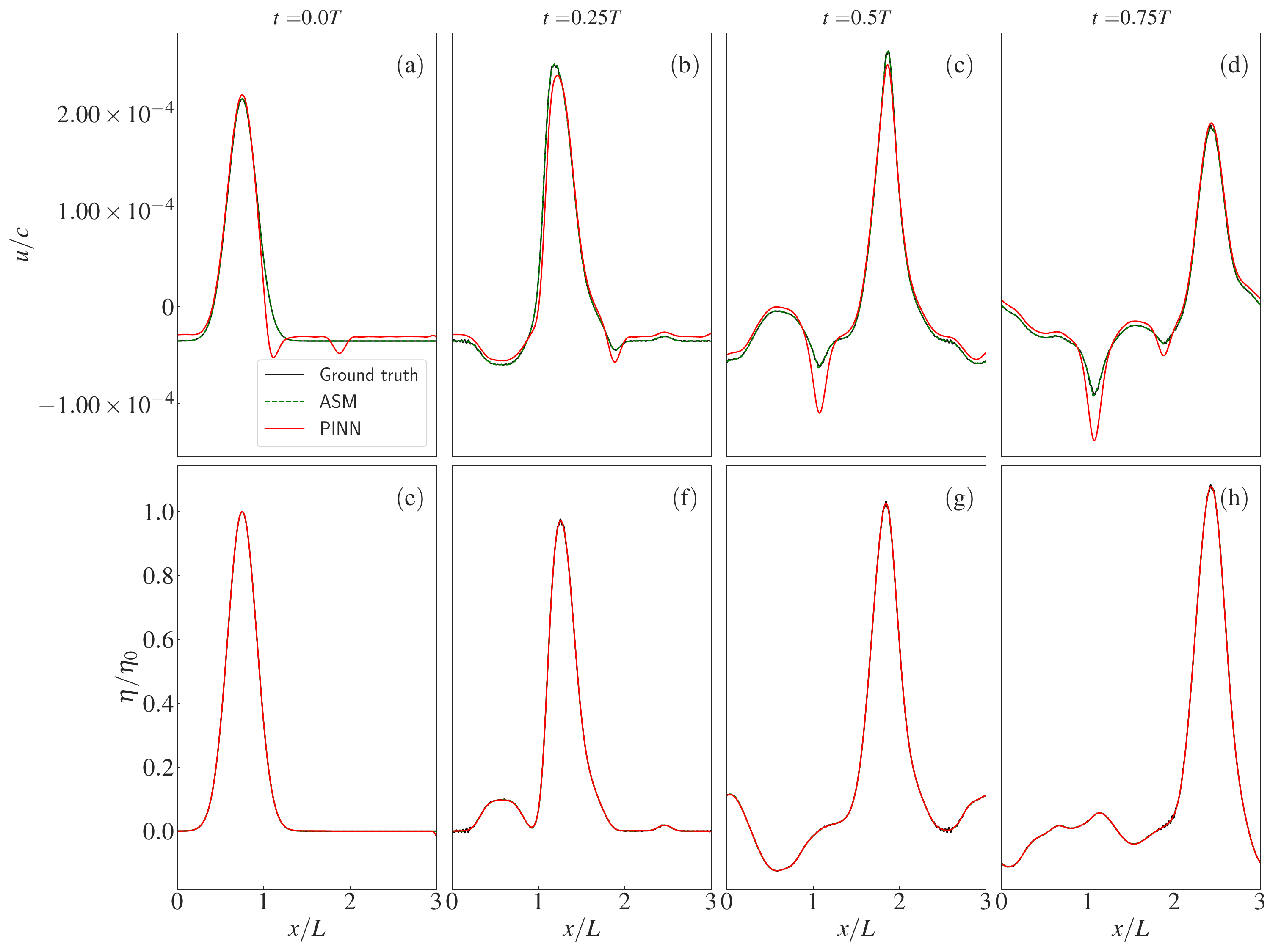}
        \caption{Case with velocity information. (a-d): True and reconstructed velocity fields $u/c$ at different times, (e-h): True and reconstructed surface fluctuations $\eta/\eta_0$;  using a separation $\delta_x=3/64$. Same legends as figure~\ref{hb_no_noise}.}
    \label{udata_uh}
\end{figure}
\clearpage

The global error produced by predictions from both methods for each amplitude of noise added to data is plotted in figure \ref{error_sweep-noise}, for (a) $\mathcal{E}_{h_b}$ and (b) $\mathcal{E}_u$. Figure \ref{error_sweep-noise}(a) shows a consistent scaling for both methods to noise in data, with a slightly more pronounced growth for $\mathcal{E}[\hat{h}_b]$. In figure~\ref{error_sweep-noise}(b) values remain relatively stable for both methods, except for an abrupt increase in $\mathcal{E}[\hat{h}_b]$ at $\epsilon/\eta_0 = 1\times10^{-1}$. \LUCAS{It is worth noting that the values of global error in figure \ref{error_sweep-noise} start close to those in the noiseless case (figure \ref{error_sweep-nx}). Here the PINN achieves smaller errors in the topography reconstruction, while the ASM does so in the velocity reconstruction.} Figure \ref{uh_noise} shows the reconstruction of fields $u/c$ and $\eta/\eta_0$ at different times in the case of noise amplitude $\epsilon/\eta_0 = 0.02$. Here the errors in the velocity reconstruction of the PINN are dominated by the appearance of non-physical artifacts near the location of the peaks.

\subsection{\REFA{Velocity measurements}}
\label{sec:uinverse}

\REFA{In this subsection, we show the results of data assimilation for the 1D flow using ASM and PINN methods, but using only measurements from field $u$ instead of the surface height $h$. In this case, the data part of the loss function changes from Eq.~\eqref{loss_general} to Eq. \eqref{loss_general_u}. It is worth mentioning that the right-hand side of Eq.~\eqref{adj_h} is now equal to zero, while the right-hand side of Eq.~\eqref{adj_u} is now equal to $\sum_{j\in \Omega_d} 2 (\tilde{u}_j - u_j)$. The distance between measurements was set to $\delta_x=3/64$ and no noise was added in this experiment. Figure \ref{udata_hb} shows ground truth (black solid line), ASM prediction (green dashed line), and PINN prediction (red solid line) for $h_b$. We observe that the assimilation is accurate. As a note, the ASM assimilation was performed without updating the initial ansatz for the initial physical fields $h(t=0)$ and $u(t=0)$ for optimization using $u$ measurements. It was found that the algorithm grew unstable when the three quantities were optimized simultaneously. The investigation of reasons for these instabilities and possible solutions are left for future work.}

Figure \ref{udata_uh} shows the results for the reconstructions of (a-d) $u$ and (e-h) $\eta$. The ground truth (black solid line), ASM prediction (green dashed line), and PINN prediction (red solid line) are plotted in each case. \LUCAS{Oscillations in the PINN method are observed in panels (a-e). These are similar to those in the reconstruction from $h$ measurements. However, the overall reconstruction remains accurate.}

\begin{figure}[t]
    \centering
         \includegraphics[width=0.8\textwidth]{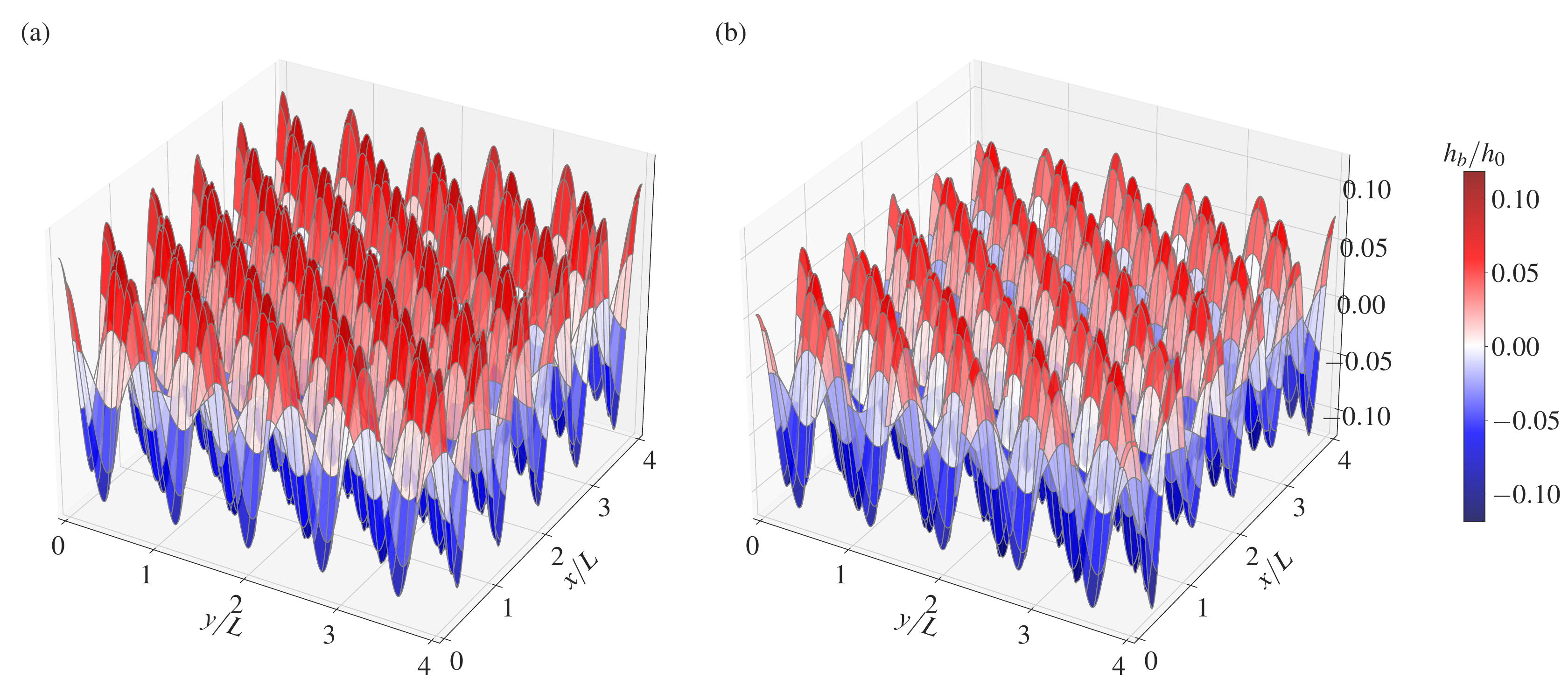}
        \caption{Surface plot of field $h_b/h_0$ for (a) ground truth and (b)  prediction using PINN method.}
    \label{cosk5_3d_hb_hbref}
\end{figure}

\begin{figure}[t]
    \centering
         \includegraphics[width=0.7\textwidth]{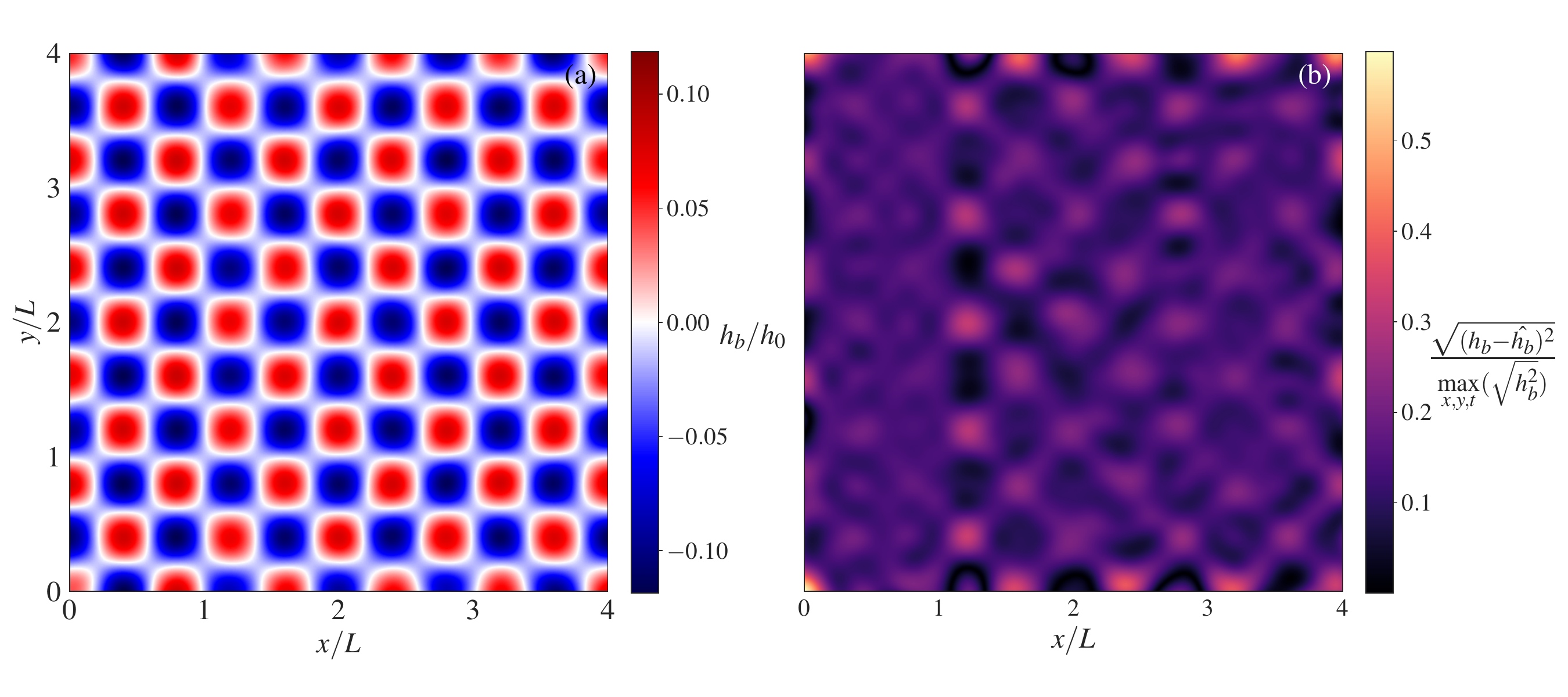}
        \caption{(a) Field $h_b/h_0$ for prediction using PINN method and (b) the error of the PINN prediction.}
    \label{cosk5_implot_hb_hberror}
\end{figure}

\begin{figure}[h!]
    \centering
         \includegraphics[width=0.7\textwidth]{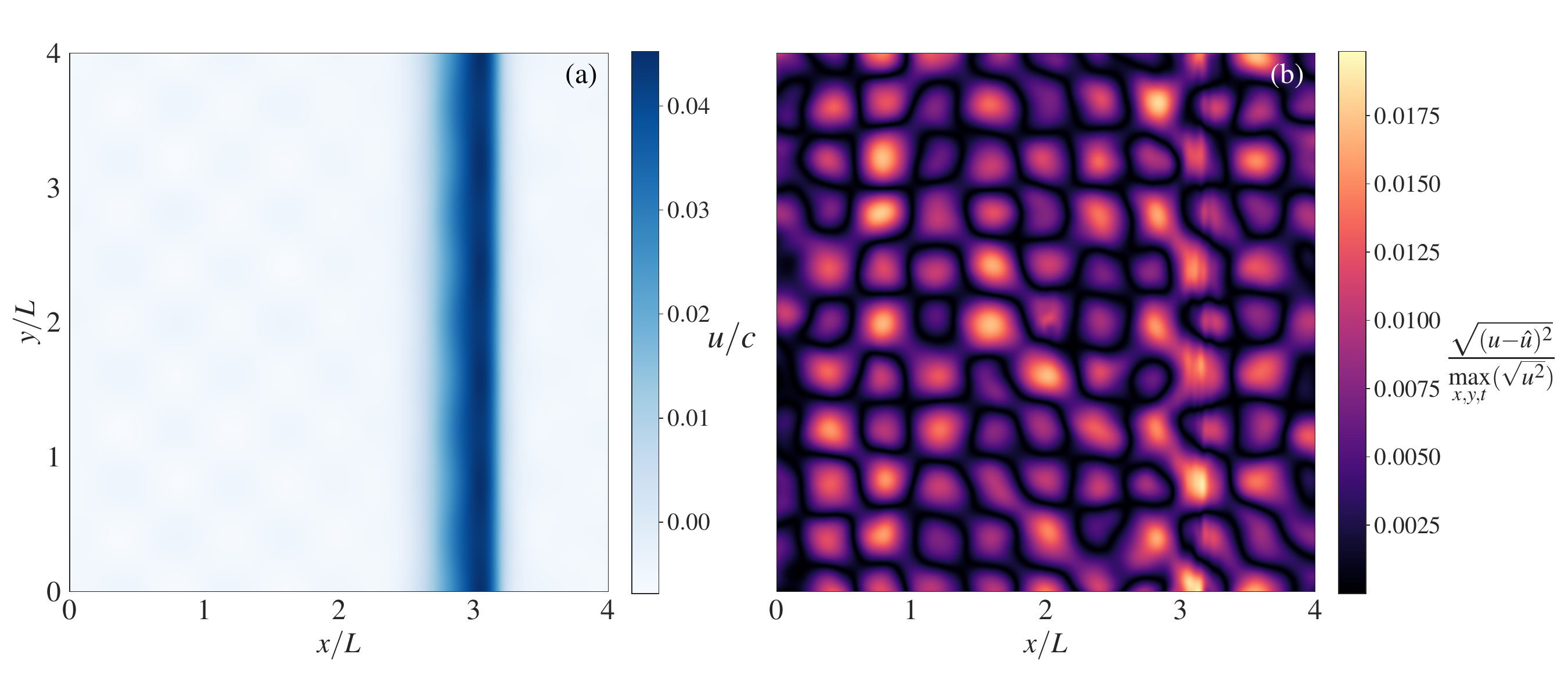}
        \caption{(a) Field $u/c$ for prediction using PINN method and (b) the error of the PINN prediction.}
    \label{cosk5_implot_u_uerror}
\end{figure}

\subsection{\REFA{Two-dimensional case}}
\label{sec:2D}

\REFA{In this subsection we discuss the results of data assimilation, using PINNs, for the 2D case. The recovery of fields $h_b$, $u$, and $v$ via assimilation of surface measurements of field $h$ took approximately 3 days running on a single Nvidia T4. In general, $h_b$ was recovered satisfactorily, as was field $u$. The smaller structures present in field $v$ proved more challenging for the networks, although this did not deter from assimilating the other fields.}


\REFA{Figure \ref{cosk5_3d_hb_hbref} shows 3D color maps of (a) the ground truth and (b) the PINN prediction for field $h_b$. The essential characteristics of the structures present in $h_b$ are recovered by the assimilation. The location of each peak, and the spatial wave number of the structures, are quite accurately reconstructed. It is seen in figure \ref{dg2d_diagram} that the bottom topography breaks the translational symmetry of the propagating wave. During the wave propagation, flow structures appear from the interaction of the gradient of $h_b$, giving it a definite 2D character.}

\begin{figure}[t]
    \centering
         \includegraphics[width=0.7\textwidth]{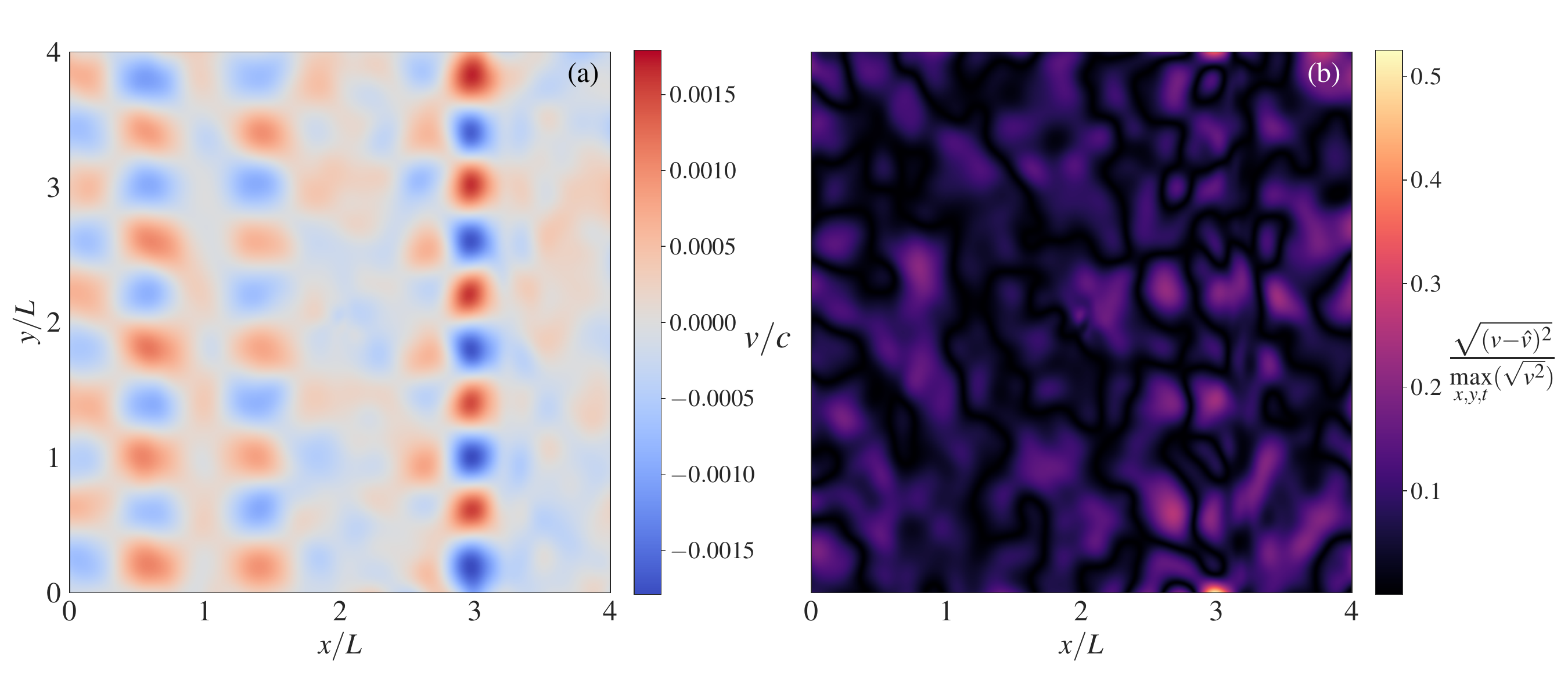}
        \caption{(a) Field $v/c$ for prediction using PINN method and (b) the error of the PINN prediction.}
    \label{cosk5_implot_v_uerror}
\end{figure}

\REFA{Figures \ref{cosk5_implot_hb_hberror}, \ref{cosk5_implot_u_uerror} and \ref{cosk5_implot_v_uerror} show 2D color maps of (a) the PINN prediction and (b) the normalized local errors of the PINN prediction, for fields $h_b$, $u$ and $v$, respectively. For time dependent $u$ and $v$ fields, the errors correspond to the same time snapshot considered for the prediction. Different time frames did not show significantly different errors for this model. We observe that the largest errors for the $h_b$ reconstruction appear to be, in general, at the location of the positive peaks of the structures. The reconstruction of field $u$ shows a similar behavior, but for negative peaks of the bottom topography structures, as well as the positive ones. In general, the error for the prediction of $u$ is of one order of magnitude less than that of $h_b$. For field $v$, we also observe a higher relative error than $u$.}

\begin{figure}[t]
    \centering
         \includegraphics[width=0.7\textwidth]{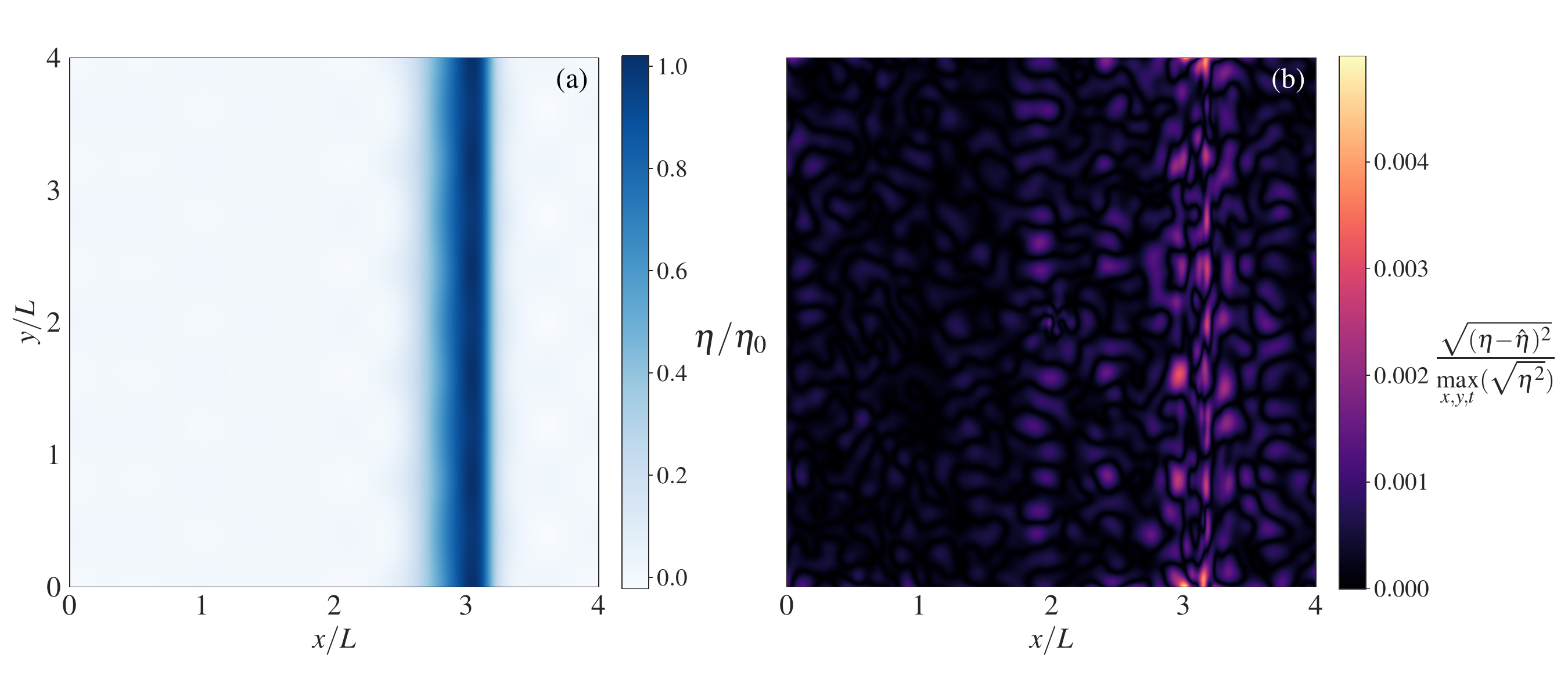}
        \caption{(a) Field $\eta/\eta_0$ for prediction using PINN method and (b) the error of the PINN prediction.}
    \label{cosk5_implot_eta_uerror}
\end{figure}

\begin{figure}[t]
    \centering
         \includegraphics[width=0.8\textwidth]{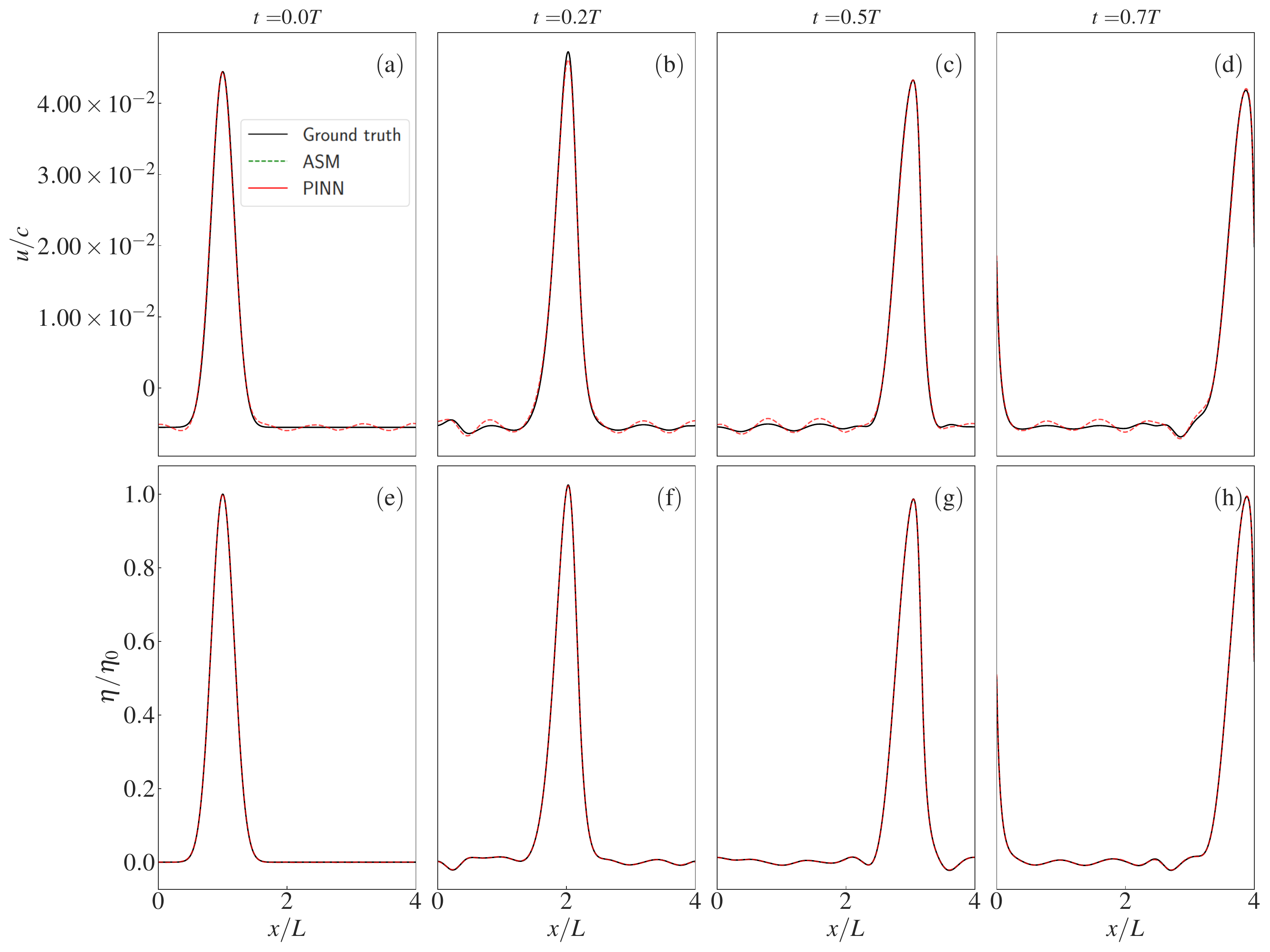}
        \caption{(a-d): True and reconstructed velocity fields $u/c$ at different times. (e-h): True and reconstructed surface fluctuations $\eta/\eta_0$. The quantities in all the panels are 1D profiles at a constant $y/L = 2$. Same legends as figure~\ref{hb_no_noise}.}
    \label{cosk5_uh}
\end{figure}

\REFA{Figure \ref{cosk5_implot_eta_uerror} shows 2D color maps of (a) the PINN prediction for field $\eta$ and (b) the error of this prediction. We observe an accurate reconstruction of field $\eta$. Figure \ref{cosk5_uh} shows different snapshots of ground truth (black solid line) and predictions (red dashed line) for fields  (a-d)  $u$ and (e-h) $\eta$, at a constant position $y_0/L = 2$, for all $x$ along the domain. The position $y_0$ corresponds to the center of the domain. We observe that both field predictions yield accurate results, with no noticeable non-physical oscillations.}

\section{Conclusions} \label{sec:conclusions}

In this work we presented two methods to estimate bottom topography using surface height data under the Shallow Water approximation. One approach is based on physics-informed neural networks (PINNs), while the other on the adjoint state method (ASM). Both are able to reconstruct the topography as well as the surface velocity field with good accuracy. \REFA{The PINN yielded a slightly better reconstruction, except for the appearance of artifacts in the velocity field.}
In particular, we were able to reconstruct structures that have about half the size of the length of the incoming wave, which should represent distances between 4 and 6km in the ocean. This \REFA{estimated accuracy is comparable with current available methods \cite{mayer-mapping-visualization}.} Results were acceptable as long as the distance between measurements  was less than half the length of the incoming wave and the noise was less than 10\%. \REFA{The PINN proved more robust than the ASM when increasing sparsity, as the tendency of the neural network to produce smooth results helped the method avoid overly jagged solutions. It is important to note that this tendency, known as spectral bias \cite{tancik_Fourier_2020}, makes it difficult to reproduce the smallest scales of the system too.} \REFA{The PINN also produced some non-physical artifacts in the velocity prediction near the location of the first peak studied, this issue could be mended by further training or by using residual-based sampling \cite{wu_Comprehensive_2022}.}
\REFA{On the implementation side, the PINN is consirably easier to implement thanks to the development of modern deep learning libraries and the overall flexibility of the framework. The ASM on the other hand requires implementing numerical solvers for both the forward and adjoint equations as well as implementing the optimization loop. Once implemented, the ASM is more computationally efficient.}

\REFA{As mentioned above, the setups studied in this work resemble that of a wave approaching shore prior to breaking or of a tsunami traveling the wider ocean. In future work we will apply both methodologies against experimental measurements and observations. Particular emphasis will be given to observed phenomena that are not captured by \LUCAS{the SW model}, such as breaking, development of smaller scales and dispersive effects, and interaction with winds. We will also study cases with waves coming from different fronts, more akin to tidal flows.}

%
%

\section*{Open Research Section}
Codes can be found at \url{https://zenodo.org/records/19244852}. The repository includes the codes used to generate the ground truth data, both the PINN and the ASM implementations of the reconstruction methods and all the scripts needed to plot the figures in the present manuscript.

\section*{Conflict of Interests Statement}
The authors have no conflicts of interest to disclose.

\acknowledgments
The authors would like to thank Pablo Cobelli for useful discussions.

\bibliography{biblio} 

\end{document}